%% file: uv_paper_arxiv.tex
\documentclass[11pt,preprint]{aastex}

\usepackage{graphicx}
\usepackage{natbib}
\usepackage{epsfig}
\usepackage{color}
\usepackage{datetime}
\usepackage{bbding}
\usepackage{lscape}
\usepackage{longtable}

\input{commands.tex}

\begin{document}

\title{The rest-frame ultraviolet properties of radio-loud broad absorption line quasars}
\author{M. A. DiPompeo\altaffilmark{1}, M. S. Brotherton\altaffilmark{1}, S. L. Cales\altaffilmark{1}, J. C. Runnoe\altaffilmark{1}}

\altaffiltext{1}{University of Wyoming, Dept. of Physics and Astronomy 3905, 1000 E. University, Laramie, WY 82071}

\begin{abstract}
We recently presented radio observations of a large sample of radio-loud broad absorption line (BAL) quasars from the SDSS and FIRST surveys, as well as a well matched sample of unabsorbed quasars, primarily to measure their radio spectral indices and estimate ensemble orientations.  Here, we analyze the SDSS spectra of these samples and compare the rest-frame ultraviolet properties of radio-loud BAL and non-BAL quasars.  Ultraviolet properties include the continuum shape, emission-line measurements of \CIV, \AlIII, \CIII, \FeII, and \MgII, and BAL properties including the balnicity index (BI), absorption index (AI), and minimum and maximum outflow velocities.  We find that radio-loud BAL quasars have similar ultraviolet properties compared to radio-loud non-BAL sources, though they do appear to have redder continua and stronger \FeII\ emission, which is consistent with what is found for radio-quiet BAL sources.  No correlations exist between outflow properties and orientation (radio spectral index), suggesting that BAL winds along any line of sight are driven by the same mechanisms.  There are also few correlations between spectral index and other properties.  We conclude that BAL outflows occur along all lines of sight with similar strengths and velocities.
\end{abstract}


\section{INTRODUCTION}
Broad absorption line (BAL) quasars represent a significant fraction of the quasar population, likely around 20\% (Knigge et al. 2008), but the reason only some quasars show BALs is still under debate.  BAL troughs in quasar spectra are the signatures of the most extreme quasar outflows.   Understanding their nature is important, as it is likely that quasar outflows can have impacts on both the evolution of the quasar (e.g. Silk \& Rees 1998, King 2003), as well as the host that harbors it, for example by regulating star formation rates (Hopkins \& Elvis 2010, Cano-Diaz et al. 2012).

Two ideas have emerged to explain the BAL phenomenon, and unfortunately they are often discussed in a mutually exclusive manner.  One is based solely on orientation, in which BAL winds are radiatively driven into a mostly equatorial outflow.  Note that ``equatorial'' here is used loosely, as in a true equatorial direction most quasars are likely obscured by dust.  Here we mean that the winds are being driven away from the symmetry axis as in Elvis (2000), for example.  In this picture all quasars have these types of outflows, and BALs are only seen when our line of sight intersects the outflow at a relatively large angle from the disk symmetry axis.  One piece of evidence used to support this view is that the emission-line properties between (radio-quiet) BAL an non-BAL quasars are quite similar (Weymann et al. 1991).  However, some differences are found, such as stronger \FeII\ emission and weaker \OIII\ lines in BAL quasars, placing them generally at one extreme of ``eigenvector 1'' (Boroson \& Green 1992, Boroson 2002).  As a class, BAL quasars more often show significant (greater than a few percent) optical polarization than their non-BAL counterparts (Ogle et al. 1999, DiPompeo et al. 2010, 2011a).  This has a natural geometrical interpretation, with polarization arising from scattering in a polar region and thus stronger from edge-on perspectives, analagous to what is seen in Seyfert type 2 galaxies (Antonucci et al. 1993).  However, other explanations for the polarization are viable as well, especially given the generally stronger reddening in BAL sources (and thus less diluting unpolarized direct continuum light), as well as the fact that polarization does not seem to correlate with orienation indicators (DiPompeo et al. 2010).  

The strongest evidence against these large inclination only explanations comes from observations at radio wavelengths.  Radio spectral index, or the steepness of the radio spectrum, is widely regarded as an orientation indicator because it depends on whether radio cores or lobes dominate the radio emission.  In jet-on, low inclination sources, core emission is relativistically boosted and overwhelms the lobe emission.  Radio cores tend to have flat spectra due to their higher optical depth.  Radio lobes, on the other hand, are optically thin and tend to have steeper spectra.  There is significant variation between sources, but when looking at large samples spectral index is a useful tool.  Studies on smaller samples found no significant difference between spectral index distributions for BALs when compared to non-BALs, indicating they have similar ranges of viewing angles (Becker et al. 2000, Montenegro-Montes et al. 2008, Fine et al. 2011).  More recently, in a larger BAL sample with a well matched non-BAL sample selected specifically for this comparison, a significant difference was found in which BAL sources had an overabundance of steep spectrum sources (DiPompeo et al. 2011b).  However, the two samples still cover a similar range in spectral index, and modeling has shown that while at the highest inclinations BALs are much more likely to be present, they very likely are also seen from small viewing angles (DiPompeo et al. 2012).  This supports the claims based on short timescale radio variability that some BALs are seen very nearly along the radio jet axis (Zhou et al. 2006, Ghosh \& Punsly 2007).

Because of these observations, the other picture that is often proposed is that BAL quasars represent an early (or rejuvinated) phase of all quasar lifetimes, in which they are still enshrouded in a cocoon of gas and dust.  The relatively recent advent of large and deep radio surveys such as FIRST (Becker et al. 1995) led to the discovery of the first radio-loud BAL quasars (Becker et al. 1997, Brotherton et al. 1998, Becker 2000), and so they have only recently begun to be studied in larger numbers.  Because of their relative rarity some have postulated that the BAL phase evolves into a radio-loud unabsorbed phase (Becker et al. 1997, Gregg et al. 2002, 2006).  Additionally, BAL quasars seem to be more compact in radio maps compared to non-BAL quasars (e.g. Becker et al. 2000), and may often have radio spectra typical of gigahertz-peaked spectrum (GPS) or compact-steep spectrum (CSS) sources (Montenegro-Montes et al. 2008).   These are generally considered to be young objects (O'Dea 1998, Fanti et al. 1995).  However, new results of larger samples of matched BAL and non-BAL quasars studied over a large range of radio frequencies suggest that the GPS/CSS fractions of BAL and non-BALs are similar (Bruni et al. 2012).  

As mentioned above, the results of DiPompeo et al. (2011b, 2012) suggest that more edge-on orientations may play a role in the presence of BALs.  But it cannot be the only factor, and BALs are seen along all lines of sight that normal quasars are seen from.  Other studies have made similar conclusions, that neither edge-on only orientations nor evolutionary sequences are sufficient to describe the BAL subclass.  For example, Allen et al. (2011) identify a redshift dependence of the BAL fraction, which seems to rule out orientation only, while Gallagher et al. (2007) find no IR excess in BAL quasars, which we would expect if all BAL sources have the large covering fractions suggested by evolution alone.

All of the quasars used in the work of DiPompeo et al. (2011b) have observed-frame optical (rest-frame ultraviolet at these redshifts) spectra from the Sloan Digital Sky Survey (SDSS).  Here we present an analysis of the ultraviolet properties of these radio-loud quasars, as well as a comparison with their radio properties, in an attempt to further develop a unified picture of BAL quasars.  We adopt the cosmology of Spergel et al. (2007) for all calculated properties, with $H_0 = 71$ km s$^{-1}$ Mpc$^{-1}$, $\Omega_M=0.27$, and $\Omega_{\Lambda} =0.73$.

\section{THE SAMPLE}
We will summarize here the sample selection, but for the full details we refer the reader to DiPompeo et al. (2011b).  The object names and redshifts are listed in columns 1 and 2 of Table~\ref{proptbl}.  The BAL sample is drawn from the catalog of Gibson et al. (2009, hereafter G09), which is built from SDSS DR5.  This catalog is cross-matched with the FIRST and NVSS (Condon et al. 1998) radio catalogs at 1.4 GHz, and sources with integrated flux densities greater than 10 mJy are selected.  No formal signal to noise cut was made, but all of the SDSS spectra were inspected by eye to ensure that there was no ambiguity in the BAL classification.  The final sample contains 74 objects.  The non-BAL sample was selected to match the BAL sample in a one-to-one fashion (for 74 more sources) in redshift, SDSS i-band magnitude, and FIRST integrated radio flux densities.  Though the G09 catalog was assembled using measurements of the SDSS DR5 catalog, all measurements performed in this analysis use the spectra from the SDSS DR7, available through the SDSS Data Archive Server (DAS).

We would also like to note here that there was in fact a non-BAL source that was mistakenly included in the BAL sample presented in DiPompeo et al. (2011b), SDSS J095327.96+322551.6.  A line-locked \CIV\ absorption doublet on the blue end of the spectrum, combined with strong \FeII\ emission near the \MgII\ line that led to the appearance of a low-ionization trough led this object to be included in the sample when it should not have.  Excluding it in no way effects the results of DiPompeo et al. (2011b) or DiPompeo et al. (2012), and it will not be included in any analysis here.  The one FeLoBAL in the sample, SDSS J155633.77+351757.3, will also be excluded in this analysis, due to the extreme absorption and associated difficulties in fitting the spectrum.

\section{MEASUREMENTS \& DERIVED PARAMETERS}
\subsection{Radio parameters}
Here we briefly summarize the radio data utilized, but again refer the reader to DiPompeo et al. (2011b) for more detailed information.  The samples were observed with the Very Large Array (VLA) or Expanded Very Large Array (EVLA), quasi-simultaneously at 4.9 and 8.4 GHz.  Integrated flux densities were measured from the radio maps, and radio spectral indices ($\alpha_{rad}$; $S_{\nu} \propto \nu^{\alpha_{rad}}$, where $S_{\nu}$ is the radio flux density and $\nu$ is the observed frequency) were measured.  There were three measures of $\alpha_{rad}$ utlilized; $\alpha_{8.4}^{4.9}$, $\alpha_{4.9}^{1.4}$, and $\alpha_{fit}$, which is a linear fit to multi-frequency data gathered from the literature in addition to the new observations.  Due to the simultaneous nature of the measurements at 4.9 and 8.4 GHz, much of the emphasis in DiPompeo et al. (2011b, 2012) was placed on that value.  However, a statistically significant difference between BAL and non-BAL $\alpha_{rad}$ distributions is seen, regardless of which $\alpha_{rad}$ is utilized.  The distributions of $\alpha_{8.4}^{4.9}$ are shown in panel (a) of Figure~\ref{propsfig}.  In this analysis, when comparing parameters to $\alpha_{rad}$, we will use both $\alpha_{8.4}^{4.9}$ and $\alpha_{fit}$.

Radio luminosities ($L_r$) are k-corrected to rest-frame 5 GHz from the FIRST survey integrated flux densities, using $\alpha_{8.4}^{4.9}$ and the SDSS redshifts.  Radio-loudness ($R^{*}$) is computed using the method of Stocke et al. (1992), using the FIRST flux densities k-corrected to 5 GHz and Johnson B magnitudes integrated from the SDSS spectra.  There are two BAL sources and two non-BAL sources for which $\alpha_{8.4}^{4.9}$ is not available, and so the average values of this spectral index for the full BAL and non-BAL samples (-0.73 and -0.36, respectively) are used. Values of $\log R^{*}$ and $\log L_r$ are listed in columns 3 and 4 of Table~\ref{proptbl}.  There is only one source that does not meet the strict definition of radio-loud ($\log R^{*} > 1.0$); the BAL quasar SDSS J110531.41+151215.9.  However, for this object $\log R^{*} = 0.98$, and so we will treat all sources as formally radio-loud.  One important note to make is that the use of B-band magnitudes can be problematic when comparing BAL and non-BAL sources.  Because BAL quasars are on average more reddened (see below), as well as the fact that the absorption troughs are often located in the B-band, differences can appear between BAL and non-BAL sources when they may in fact be intrinsically the same.  However, the $\log R^{*}$ distributions of the two samples here are indistinguishable and the samples are matched in the SDSS i-band (where reddening and BAL features are less of an issue), and so we are comfortable using the traditional definition of $R^{*}$ using B-band magnitudes.

\subsection{Spectral fits}
The SDSS DR7 spectra were corrected to rest-frame wavelengths (flux densities were preserved as observed-frame), corrected for Galactic extinction using the dust maps of Schlegel et al. (1998), and then fit using the \textsc{specfit} task (Kriss 1994).  The components used in the fits were a power-law continuum ($F_{\lambda} \propto \lambda^{- \alpha_{uv}}$), the I Zw 1 UV iron emission template (Vestergaard \& Wilkes 2001), and two Gaussians each for the emission lines of \CIV\ $\lambda$1549 \AA\ (for the non-BAL sample only), \AlIII\ $\lambda$1857 \AA, \CIII\ $\lambda$1909 \AA, and \MgII\ $\lambda$ 2799 \AA.  No attempt to de-blend \SiIII\ $\lambda$ 1892 \AA\ from the \CIII\ line was made.  One of the Gaussian components for each line was allowed to be narrow enough to be considered a traditional narrow line (FWHM $<$ 1000 km/s), but in all cases both Gaussians were broad in the best fits.  In some cases additional Gaussians were used to approximate the \HeII\ $\lambda$ 1640 \AA\ and \OIIIsemi\ $\lambda$ 1663 \AA\ emission on the red side if \CIV, so that the fitting procedure did not try to use one or both components of \CIV\  to fit the ``shelf'' created by these other emission lines, artificially increasing the \CIV\ parameters.

We aimed to be as consistent as possible with the regions included in the fits, but the large redshift range limits our ability to do this.  In the non-BAL sample, fits were performed beginning at 1450\AA\ (just blueward of the \CIV\ emission), or as far blue as the redshift would allow if 1450\AA\ was not in the spectral window.  Fits extended up to 3000\AA\ when possible.  In cases where \MgII\ was redshifted out of the spectrum, fits were performed out to 2500\AA, and if that was not available, then 2250\AA\ was used.  The BAL sample was fit in a similar way, except the \CIV\ emission was excluded because of the uncertainty due to absorption and the fits began at 1600 \AA.

Individual spectra were inspected, and regions of bad data or other absorption (due to low-ionization absorption or intervening systems) were excluded from the fit regions.  Several sets of fits were run, using the same upper and lower limits for parameters that were used in the prescriptions of Ganguly et al. (2007).  This includes allowing the \FeII\ template to be broadened or narrowed, as well as have a velocity shift relative to the other lines; both of these were freely varying parameters in the fits.  Separate sets of fits were performed both tying the peak wavelengths of the two Gaussian components of each line together and allowing them to vary independently.  Once all fits were run, those with the lowest $\chi^{2}$ for each object were chosen as the final fits, though all fits were inspected to ensure that these ones were indeed the best model of the data.  In some cases it was necessary to go back and modify the fits interactively, and in situations where a parameter could not be satisfactorily fit, it was thrown out.  

\begin{figure}
\centering
 \includegraphics[width=3.25in]{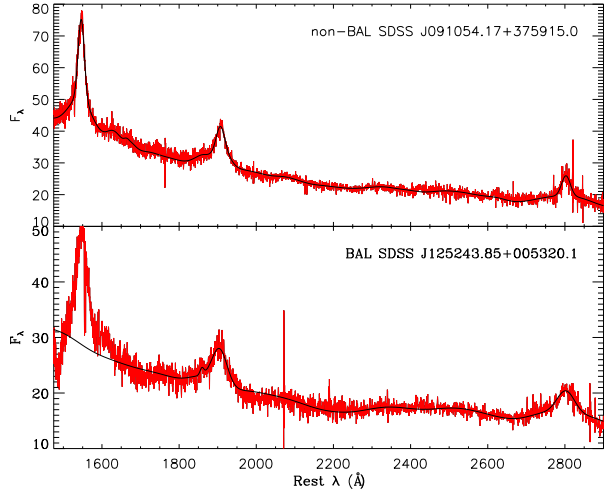}
 \caption{Two examples of our spectral fits, for a non-BAL (top) and BAL (bottom).  Recall that fits on the BAL sample do not include the \CIV\ line due to uncertainties from absorption, and so the fits are performed long ward of 1600 \AA.  Some systematic differences from the \FeII\ template used are seen in many objects, where the fit is slightly low from about 2400-2500 \AA\, and slightly high just redward of the \CIII\ emission line.  Flux is in units of $10^{-17}$ erg/s/cm$^2$/\AA.}
 \label{fitsfig}
\end{figure}

We point out that for sources with particularly strong (or narrow) \FeII\ emission, it was often difficult to get the Vestergaard \& Wilkes (2001) template to fit well in all spectral regions (in particular around 2400-2500 \AA\ and immediately to the red side of \CIII), indicating that the template line ratios may not be ideal for all sources.  However, it generally would fit well around the emission lines, with the exception of the red side of \CIII, and the \FeII\ normalization factor is still a useful diagnostic of \FeII\ strength even with these shortcomings.  There are some cases where separating the \CIII\ and \AlIII\ emission is difficult, particularly when signal-to-noise is lower, as it is harder to tell if \AlIII\ has a distinct peak or if it is completely blended with \CIII.  However, the fact that our results do not change much when lower SNR spectra are omitted indicates that our fits are robust.  In general the fits are well within the noise of the actual spectra, with few systematic discrepancies.  Examples of our fits to a BAL and non-BAL source are shown in Figure~\ref{fitsfig}.  The parameters of the continuum fit and \FeII\ normalization (relative to the I Zw 1 template) are shown in columns 2, 3, and 4 of Table~\ref{lineproptbl}.  Distributions of the power-law slopes $\alpha_{uv}$ and \FeII\ scaling factors for the BAL and non-BAL samples are shown in panels (b) and (c) of Figure~\ref{propsfig}.  These are shown because they are two of the most statistically different distributions, as discussed in Section 4.1.

Since the two Gaussian components used for the emission lines are not likely to be from different physical regions (as both components are always broad), we combine them into a single, total line profile.  This was done for each emission line numerically.  We then measure the centroid, FWHM, and EW of the total line profiles.  Finally, the line centers of \CIV, \CIII, and \MgII\ are converted into blueshifts (in km/s).  For consistency with the measurements of Richards et al. (2011), and to test some of their ideas later, we use the modified SDSS redshifts of Hewett \& Wild (2010) in this calculation, as opposed to the original SDSS redshifts used in the other calculations.  The final EWs, FWHMs, and blueshifts of each line are presented in columns 5-15 of Table~\ref{lineproptbl}.

Errors on the fit parameters were estimated using the following method.  Spectra with five representative signal-to-noise ratios (approximately, 5, 10, 15, 25, and 35 in the i-band, as reported in the SDSS headers) were chosen.  We then added random Gaussian noise to the original fits of these spectra consistent with their SNR and performed fits to these synthesized spectra.  Individual line components were combined in the manner described above, and the process was repeated fifty times for each of the five spectra.  We then used the standard deviation of the distributions of each parameter as an estimate of the errors.  In order to extrapolate these errors to other spectra, we fit a power-law to the relationship between the errors from the 5 representative spectra (normalized by their values from the initial fits) and SNR.  The use of a power-law relationship was chosen by visual inspection of the shape of the normalized error-SNR data.  These fits allow us to rescale the errors to other spectra with different SNRs and parameter values without having to fit 50 synthesized spectra for each object, saving significant computing time.  We note that using this method may under or over-estimate errors for some parameters in individual spectra, but overall they provide a good representation of the errors.

A few points should be made about this error estimation technique.  First, since the mock spectra were generated from the model fits, it is possible for the errors to be underestimated as technically the mock spectra can be fit exactly while the real spectra will almost always have some residuals.  However, since generally our fits are within the noise of the spectra (see for example Figure~\ref{fitsfig}), this effect is minimal.  Additionally, since errors on individual objects were extrapolated from just a few representative spectra, issues with the accuracy of the errors are likely dominated by that step.  It is important to also note that this method does not fully take into the account the covariance of fit parameters.  However, the fact that Gaussians were combined for each of the 50 fits (i.e., a Gaussian from one fit wasn't combined with a second Gaussian from a different fit), this is accounted for at some level.  While the \textsc{specfit} task does calculate errors on individual fit components using full covariance matrices, there is no straightforward way of combining these errors on individual Gaussians for an error on the final line profile, particularly the FWHM.  Additionally, the covariance matrix method does not take into account the noise in the spectra, which is a significant source of errors.  The exact way in which errors are estimated will not effect our final results, and thus we feel that this method is sufficient.

\subsection{Bolometric luminosity \& black hole mass}
Using the fits described above, we next calculate luminosities and virial black hole masses.  Bolometric luminosities ($L_{bol}$) are calculated using a bolometric correction at 2500\AA, derived in the same way and with the same sample as the correction at 3000\AA\ in Runnoe et al. (2012).  Though not published there, the correction at this wavelength was readily available.  Since 2500\AA\ is covered in significantly more of our spectra than 3000\AA, less extrapolation of the continuum fits is needed there and so its use is preferred.  The correction at this wavelength is:
\begin{equation}
\log L_{bol} = 2.653 + 0.957 \log (2500L_{2500}).
\end{equation}
Bolometric luminosities are listed in column 5 of Table~\ref{proptbl}.

We also estimate virial black hole masses ($M_{bh}$) when the \MgII\ line is in the spectral window.  Though the \CIV\ line can also be used to calculate $M_{bh}$, since we did not attempt to fit \CIV\ in the BAL sample we do not use this method here.  We follow the prescription of Vestergaard \& Osmer (2009), using their luminosity calibration at 2100\AA:
\begin{equation}
M_{bh} = 10^{6.79} \left( \frac{\rmn{MgII\ FWHM}}{1000\ \rmn{km/s}} \right)^{2} \left( \frac{2100 L_{2100}}{10^{44}\ \rmn{erg/s}} \right) ^{0.5}
\end{equation}
Black hole masses are listed in column 6 of Table~\ref{proptbl}.

Finally, we use $M_{bh}$ to calculate the Eddington fraction ($L_{bol}/L_{edd} \equiv F_{edd}$) of the sources, using the equation of Peterson (2003) for the Eddington luminosity:
\begin{equation}
L_{edd} = 1.51 \times 10^{38}\ M_{bh}
\end{equation}
Eddington fractions are listed in column 7 of Table~\ref{proptbl}.  The distributions of $L_{bol}$, $M_{bh}$, and $F_{edd}$ are shown in panels (d), (e), and (f) of Figure~\ref{propsfig}.

\subsection{Absorption properties}
In order to explore the absorption properties of BALs as a function of viewing angle and other properties, we make several measurements of the \CIV\ absorption features.  Though we agree with the idea that many BAL definitions and measurements are somewhat arbitrary (e.g. Ganguly \& Brotherton 2008), they can still be useful in searching for correlations between outflow and quasar properties.  Using the SuperMongo code made available by Hall et al. (2002), we first measure the traditional ``BALnicity index''  ($BI$; Weymann et al. 1991):
\begin{equation}
 BI = \int_{3,000}^{25,000} \left(1-\frac{f(v)}{0.9} \right) C dv
\end{equation}
where $f(v)$ is the continuum normalized flux density as a function of velocity (in km/s), and $C$ is set to 0 except for in regions where the normalized flux density is below 0.9 for 2000 km/s continuously, where it is equal to one.  Values of $BI$ are given in column 8 of Table~\ref{proptbl}.  The continuum used for normalization is not technically the same as the power-law fit discussed above.  Because fits were only performed beginning on the red side of the \CIV\ emission line in the BAL sample, in some cases the continuum fit does not extrapolate well to the BAL region.  This is particularly true for sources in which reddening causes a turnover in the spectrum around the \CIV\ region.  Instead, we fit a 3rd-5th order polynomial to all regions of the spectrum without emission or absorption features, similar to the method of Hall et al. (2002).  Over the relatively small wavelength region where \CIV\ BAL features are measured, the differences between a power-law and polynomial fit are generally small.  For consistency we have measured the absorption properties of all sources using the polynomial fit, but in cases where the power-law fit appears to extend through the BAL region fairly well we have checked that the power-law and polynomial fits give consistent answers.

Because $BI$ was originally defined to only identify the strongest BAL features, there are often obviously strongly (and intrinsically) absorbed systems with $BI=0$, which we see in many of our sources.  Because of this, we also measure the ``absorption index'' ($AI$), which is similar to $BI$ but allows for lower velocity (and sometimes narrower, see below) absorption:
\begin{equation}
 AI = \int_{0}^{25,000} \left(1-\frac{f(v)}{0.9} \right) C' dv
\end{equation}
For consistency with the measurements of G09, we use a condition of continuous absorption for 2000 km/s when assigning a value to $C'$.  Unlike for $BI$, this is not necessarily built into the standard definition of the parameter, and the user can define any continuous absorption condition.  However, in a significant number of cases, using this condition results in a measurement of $AI=0$, despite a non-zero measurement by G09.  This could be for several reasons.  One, G09 smooth their spectra before measurement, while we do not, but this is only an issue in a small fraction of objects.  Two, our continuum placements may differ slightly, and so the width of the absorption below the continuum may fall below 2000 km/s.  This is the most common difference, and occurs often when there is significant absorption contained within the \CIV\ emission line, causing only the narrower parts of the absorption to fall below the continuum.  For sources which we measure $AI=0$ using these conventions, we relax the continuous absorption condition to 800 km/s, which allows us to find $AI > 0$ for all objects.  These measurements highlight some of the difficulties in measuring BAL properties accurately, as well as the difficulties in defining them such that true BALs aren't missed and false positives are avoided.  Values of $AI$ are tabulated in column 9 of Table~\ref{proptbl}.  In both $BI$ and $AI$ measurements, $v=0$ is defined to be at 1549.06\AA\ and  integration is only carried out to 25,000 km/s because this is the velocity at which the \SiIV\ region begins.  This and the fact that in some cases regions of possible absorption lie outside the spectral coverage (for $z<1.6$) means that some BAL parameters are only lower limits.

Finally, we measure the maximum and minimum velocities ($v_{max}$ and $v_{min}$, respectively) of the BAL outflow, also relative to $v=0$ at 1549.06\AA.  The values of $v_{max}$ are taken from the $AI$ integration, and so are where the absorption finally rises above 90\% of the continuum.  $v_{min}$ is measured more subjectively, because the minimum velocity may still be located on the emission line.  Therefore, using the location where the flux density first drops below the continuum can give values of $v_{min}$ that are biased too high.  Each of our spectra were inspected and the location of $v_{min}$ was estimated by eye.  These velocities are given in columns 10 and 11 of Table~\ref{proptbl}.  Negative values of $v_{min}$ indicate that the absorption begins to the red of 1549.06\AA. 

Estimating errors on outflow properties is difficult.  While formal errors for $AI$ and $BI$ are relatively straight forward to calculate (indeed they are reported by the Hall et al. (2002) code) using the SDSS error spectra, they are not generally an accurate measure of the true errors.  The real errors are dominated by other factors, particularly continuum placement.  Formal errors on $AI$ are on average only about 2\%, with the lowest error being 0.1\% and the highest being 10\%.  However, if we take a few example spectra and fit several different polynomial continua (judging by eye what a ``maximum'' and ``minimum'' continuum might be), the variance in these measurements is always larger than the formal error.  In some cases this error can be nearly 50\%, though this a function of SNR and the continuous absorption criteria (narrower absorption is generally more sensitive to continuum placement).  Errors on $AI$ and $BI$ are likely around 5-10\% on average.  Errors on $v_{max}$ due to continue placement are generally a bit less as in most cases the velocity where $AI$ integration ends does not change significantly as the continuum changes.  Errors on $v_{min}$ are difficult to estimate, due to the subjective way in which they are measured.  However, for both $v_{min}$ and $v_{max}$ redshift errors can be important.  The average redshift error for the BAL sample here is 0.0017, or a velocity of about 500 km/s, and so this can be taken as an average error on velocities.  However, redshift errors are random and thus shouldn't significantly affect the correlation tests as the errors should average out.

\subsection{Comparisons with other measurements} 
Being drawn from SDSS, many of the measurements used here have been independently made by others, providing a good opportunity for comparisons that may prove useful in analyzing various methods.  First, for the BAL sample, measurements of the \CIV\ BAL properties are presented in the G09 catalog that this BAL sample is derived from.  The spectra of all of the objects in both of our samples have been fit by Shen et al. (2011, hereafter S11), and fit parameters from the SDSS pipeline are available as well.  The parameters of the SDSS fits are taken from the publicly available table of results in Richards et al. (2011).  Although there is 100\% overlap between these sources, there are cases where some measurements are made of a parameter by us and not by S11, or vice-versa.  This can be for several reasons.  For example, S11 does not measure the \CIV\ profile when it is on the edge of the spectrum, while we include measurements in these cases where our visual inspection indicates that the fit looks reasonable.  As mentioned above, in come cases fit parameters (particularly of \MgII) were thrown out in our procedures due to things like narrow absorption, BAL troughs, or simply via visual inspection if we did not feel the fit was accurate.  Many of these objects have fit parameters presented by S11.

We refer the reader to S11 for details of their fitting procedures, but we highlight a few of the main differences between this work and theirs.  While both of our methods use a power-law continuum and similar \FeII\ template, a key difference is that these components are fit to all spectral regions simultaneously in our fits, while S11 fit the regions around the emission lines individually.  So while our continuum+iron is restricted to be consistent across the whole spectrum, the S11 method allows for it to be different around the different emission lines.  Also, S11 use up to 3 Gaussians to fit each of the \CIV\ and \MgII\ broad emission (plus an additional narrow component for \MgII), while we use two for each.  In contrast, the SDSS pipeline fits a simple polynomial to the continuum, with no \FeII\ estimation, and single Gaussians to the emission lines.

First, we compare emission-line equivalent widths for \CIV\ and \MgII\ emission, shown in the top rows of Figures~\ref{civcomp} and \ref{mgiicomp}, respectively.  The reader will notice in these figures a few objects included in some panels but not others, for the reasons mentioned above.  The SDSS pipeline EW values are systematically lower than the measurements presented here and those in S11, worse so for \MgII.  This is likely due to the fact that the pipeline fits a single Gaussian profile to the lines, when in general AGN broad lines require multiple Gaussians for accurate fits, and so the single profile misses some of the line flux and leads to a lower EW.  The effect isn't simply systematic either, as the disagreement generally worsens for higher EW.  Analysis using these SDSS fits, as in Richards et al. (2011), should be treated with caution.  The SDSS pipeline often does a very poor job fitting the generally noisier \MgII\ line.  In several instances it fits noise spikes rather than the line itself, and very small EW values are returned.  In at least one case, it even fits a narrow absorption system and returns a negative value of EW.

For the \CIV\ line, our EW measurements generally agree quite well with those in S11, though on average their values may be slightly larger than ours (by about 10\%).  As mentioned in their analysis, their values of \CIV\ EW may be slightly overestimated because they do not subtract \FeII\ emission in that region.  The agreement for \MgII\  EW is not as good.  There is significantly more scatter, and on average our values tend to be higher (by about 20\%).  Since \FeII\ emission is much more significant in this region, it is likely that the difference is due to the way in which \FeII\ is fit.  Again, because we require that \FeII\ be fit simultaneously in all spectral regions, we may include less \FeII\ in this region on average than S11, and in turn include a more significant broad \MgII\ component to make up for it.  The two largest outliers in the \MgII\ FWHM comparison are good examples of this, as discussed below.  Differences between the compared values seem independent of whether the source is a BAL or unabsorbed quasar.

Better agreement between the SDSS measurements and ours (and S11 as well) are seen for the FWHMs, shown in the bottom rows of Figures~\ref{civcomp} and \ref{mgiicomp}.  Again, the agreement between the 3 methods is better for \CIV\ than for \MgII.  On average, we find that our values for the \CIV\ FWHM are slightly lower than in S11 (by 4\%), but the opposite is true for the \MgII\ FWHM (by 16\%).  However, with the exception of a few outliers, our values and those of S11 agree fairly well.  Again, both BAL and non-BAL quasars are just as likely to have differing values.

The largest outlier in the \CIV\ FWHM comparison is due to our inclusion of a narrower component that brings down the FWHM.  Via visual inspection it is hard to say which measurement is correct due to the noise level, but there does appear to be a narrower component to the line that may be missed by S11.  The two largest outliers in the \MgII\ FWHM comparison are illustrative of the issues with the \FeII\ contribution mentioned above.  Both sources are quite noisy in the \MgII\ region, and for both sources there appears to be a blue asymmetry to the line which could be due to \FeII\ emission or a slightly offset broader \MgII\ component.  While it is difficult to tell these scenarios apart, the fact that \FeII\ appears well fit in other regions leads us to believe that the emission is really due to \MgII.  Thus, we measure much larger FWHMs than S11.  In the case of the BAL quasar in this region, which has some narrow absorption on the line complicating the issue, it appears to us very unlikely that the width of the line is less than 4000 km/s as suggested by S11.

We next compare the values of $M_{bh}$ and $L_{bol}$ calculated in this work and in S11, shown in Figure~\ref{masslumcomp}.  It is important to note that the two values of $L_{bol}$ are computed using different corrections (S11 uses the correction of Richards et al. 2006), but nevertheless there is overall good agreement.  The largest outliers here seem to be BAL sources, but it is unclear why this should be the case.  The overall good agreement between our measures of \MgII\ FWHM and continuum luminosities is reflected in the agreement between our $M_{bh}$ calculations (the values presented here from S11 use the same scaling relationship used in this work).  However, the tendency for our \MgII\ FWHM to be larger can be seen, with more of the points falling below the dashed line.

Finally, we compare the \CIV\ BAL properties with those measured in G09, shown in Figure~\ref{outflowcomp}.  The left panel shows the comparison between $BI$ (pluses) and $AI$ (which G09 call $BI_{0}$; squares) measurements.  Overall the agreement is fairly good, considering the difficulties measuring these parameters consistently.  Our $AI$ values tend to be slightly larger, and BI values slightly smaller than those of G09.  However, there are numerous cases in which we measure an $BI$ of 0 and G09 does not, or vice versa.  This would be the case for $AI$ as well, if we only used the measurements using the 2000 km/s continuous absorption requirement.  Recall that we specifically avoid this by using 800 km/s if 2000 km/s returns a value of 0.  Again this decision was made to avoid the presence of false positives or false negatives in the $AI$ measurement, which can clearly be a fine line.  This generally produces good agreement in our results and those of G09. 

The middle panel of Figure~\ref{outflowcomp} shows a comparison of $v_{max}$, and we agree quite well in most cases, again despite the different integration constraint on $AI$ for some sources.  The few big outliers again highlight the difficulties of making these measurements in some sources.  Via visual inspection we see absorption we believe is real and make sure it is picked up in the integration for two sources (below the line) that G09 do not agree with.  The opposite is true for the source at the top left.  

The panel on the right of Figure~\ref{outflowcomp} shows a comparison of $v_{min}$.  Again many sources show excellent agreement, but in this case the values from G09 are often higher than those measured here.  This is due to a slightly different approach in making the measurement.  In G09, $v_{min}$ is measured where the absorption first drops below 90\% of the continuum flux density.  This will miss the lower velocity absorption that is still high up on the \CIV\ emission line.  When making our more subjective measurements, $v_{min}$ is marked at the point where it is clear that absorption is beginning, even if the flux density there is still above the continuum, which leads to lower values.  Because there are significant differences between our \CIV\ BAL measurements and those of G09, we performed correlation tests (see below) using both sets of values.

\section{ANALYSIS \& RESULTS}

\subsection{BAL versus non-BAL properties}
One key purpose of this work is to search for differences in the average properties of radio-loud BAL and non-BAL sources.  These comparisons are quantified in Table~\ref{balnbalcomptbl}, where we present the statistics of our measured/calculated parameters for the BAL and non-BAL samples.  Listed in the table are the mean ($\mu$), standard deviation ($\sigma$), and number ($n$) of sources with a given measurement for the BAL sample (columns 2-4) and non-BAL sample (columns 5-7).  We also compare the distributions of these parameters using both a Kolmogorov-Smirnov (KS) test and a Wilcoxon rank-sum (RS) test.  The final 4 columns of Table~\ref{balnbalcomptbl} give the KS statistic ($D$) with the associated probability that the BAL and non-BAL distributions are drawn from the same parent population ($P_{KS}$), and the RS statistic ($Z$) with the associated probability that the samples have the same mean ($P_{RS}$).  We define a significant difference between the two samples as having a value of $P_{KS}$ or $P_{RS}$ of $0.01$ or less, and these values are typeset in bold.  All of the tests were also performed excluding LoBALs from the BAL sample, as well as excluding objects with a $SNR <  6$ in the SDSS i-band.  The results were very similar, and so only the results including all available objects are presented.  

\begin{figure}
\centering
 \includegraphics[width=6.5in]{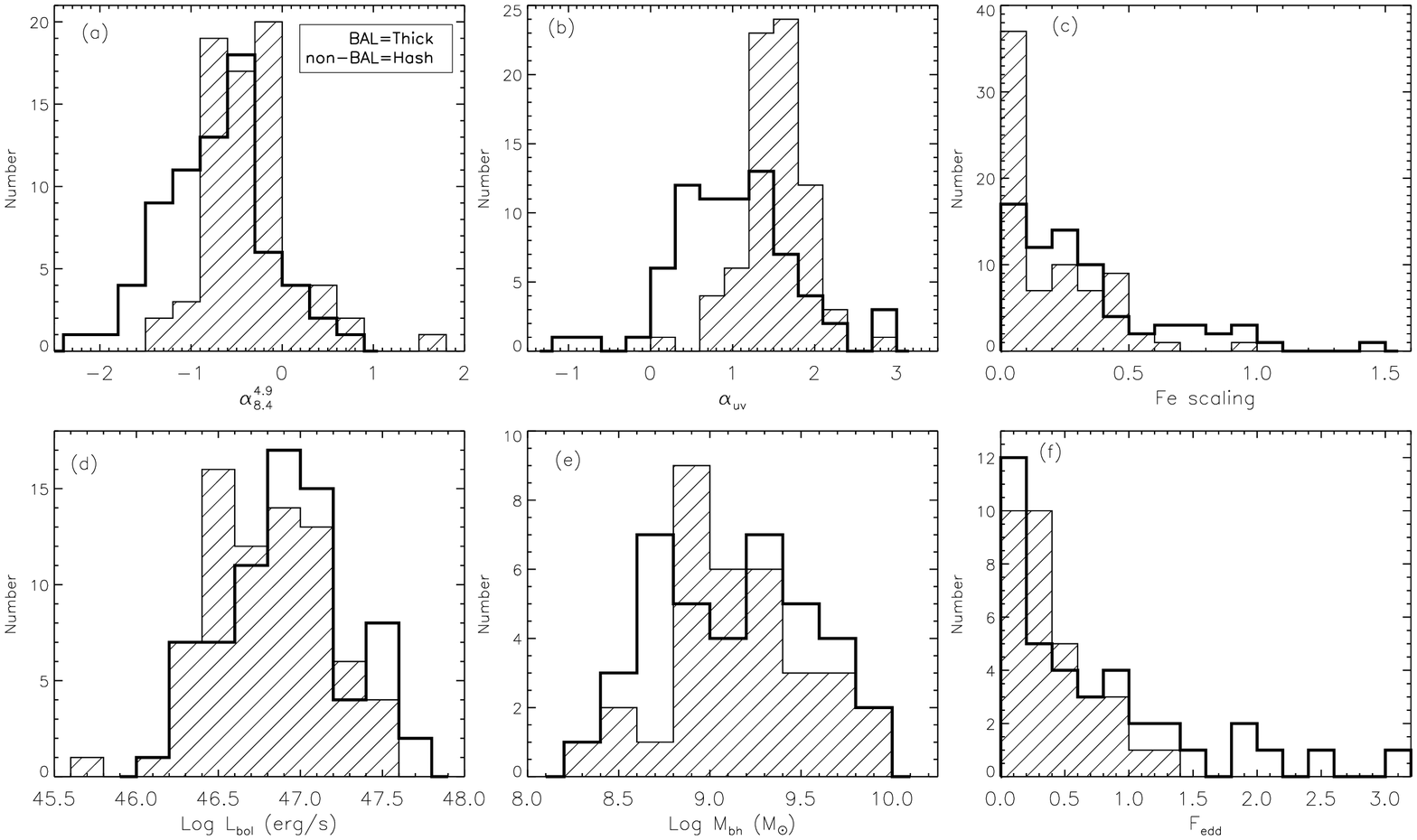}
 \caption{The distributions of properties of the sample; all plots follow the legend in the first panel.  (a) The radio spectral index $\alpha_{8.4}^{4.9}$, reproduced from DiPompeo et al. (2011b).  (b) The ultraviolet spectral index $\alpha_{uv}$.  (c) The scaling factor of the Vestergaard \& Wilkes (2001) \FeII\ template.  (d) The bolometric luminosity.  (e) The black hole mass, calculated from the FWHM of \MgII\ using the method of Vestergaard \& Osmer (2009).  (f) The Eddington fraction.}
 \label{propsfig}
\end{figure}

\begin{figure}
\centering
 \includegraphics[width=6in]{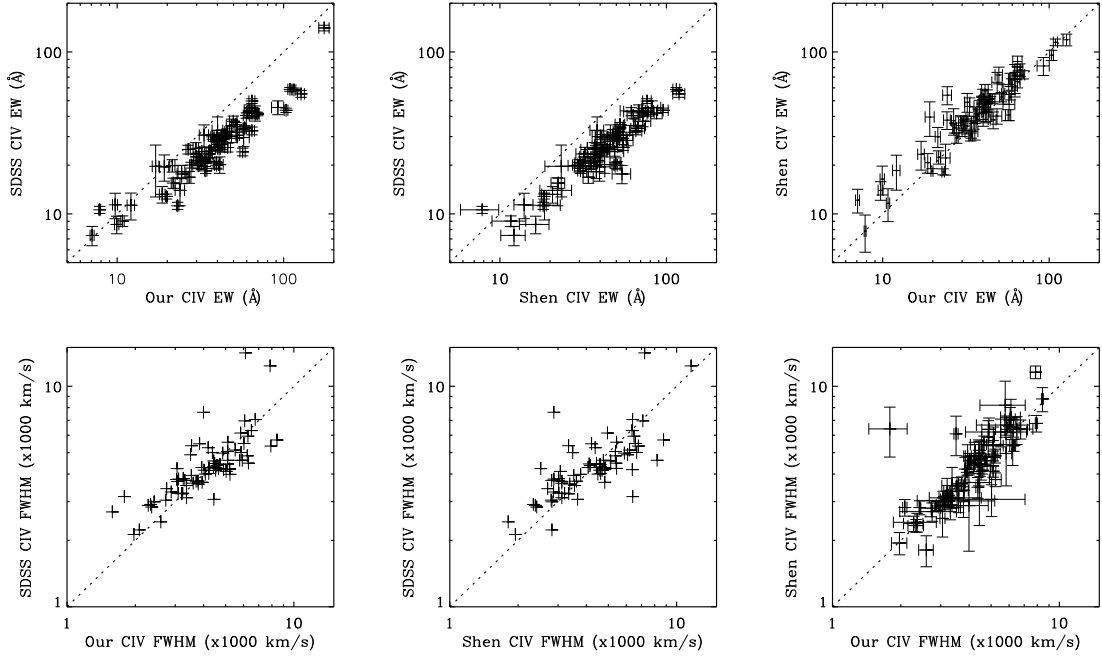}
 \caption{Comparison of the \CIV\ EW (top row) and FWHM (bottom row) measurements from this work, the SDSS pipeline, and the fits of Shen et al. (2011).  Note that in some cases a measurement was made for one object by one of us, but not by others.  Reasons for this vary, but could be due to intervening absorption, BAL features, or noisy data (for example, there are several sources where we did not feel a reasonable fit could be made to a particular line, but the measurement was included in the catalogs of others).  Therefore, some sources appear in one panel but not others.  We do not show the errors on the SDSS FWHM plots, because we do not calculate them after converting the SDSS Gaussian fit standard deviation into a FWHM.  Dotted lines indicate where the values are equal.}
 \label{civcomp}
\end{figure}

\begin{figure}
\centering
 \includegraphics[width=6in]{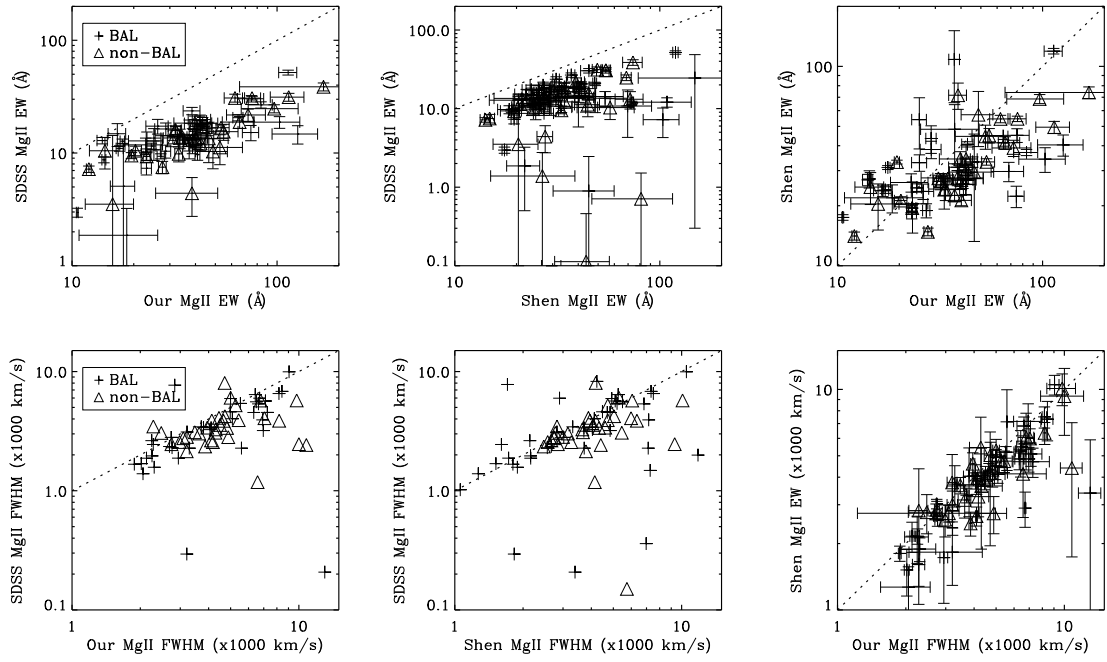}
 \caption{The same as Figure~\ref{civcomp}, but for \MgII\ emission measurements.}
 \label{mgiicomp}
\end{figure}

\begin{figure}
\centering
 \includegraphics[width=2.75in]{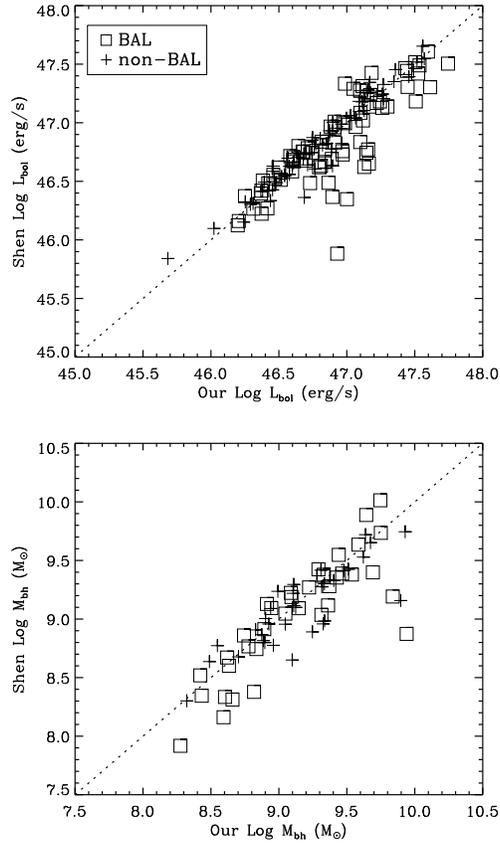}
 \caption{Comparison of our derived $L_{bol}$ and $M_{bh}$ with those from Shen et al. (2011).  Despite our disagreement in some of the individual fit parameters, there is generally good agreement in these calculated properties.}
 \label{masslumcomp}
\end{figure}

\begin{figure}
\centering
 \includegraphics[width=6.5in]{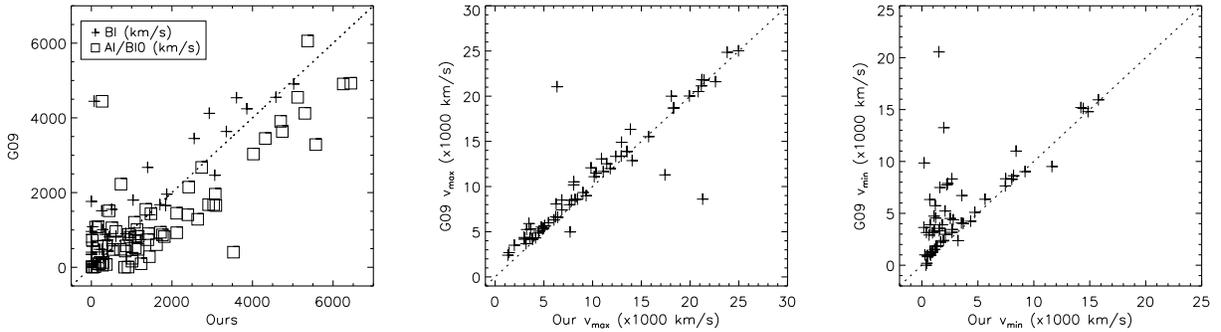}
 \caption{Comparison of our \CIV\ BAL measurements with those reported by Gibson et al. (2009).  The scatter in $BI$ and $AI$ highlight the difficulties in making these measurements consistently.  Excellent agreement is seen in $v_{max}$, except for a few sources in which we identify some high velocity absorption that G09 does not or vice-versa, again showing the difficulties of these measurements.  The tendency for our values of $v_{min}$ to be lower than G09 is due to the way in which it was measured.  G09 uses the start of $AI$ integration to determine $v_{min}$, while we measure it subjectively to include low velocity absorption on the absorption line and thus well above the continuum.}
 \label{outflowcomp}
\end{figure}

Overall, the BAL and non-BAL samples are remarkably similar when using these statistical comparisons.  The significant differences between BAL and non-BAL quasars are in $\alpha_{uv}$ (Figure~\ref{propsfig} b), \FeII\ strength (Figure~\ref{propsfig} c), \AlIII\ EW, and possibly in \CIII\ FWHM (only the RS test meets our $P$ value cutoff).  It is possible that this slight difference in \CIII\ is actually an indicator of stronger \SiIII\ in BAL sources, as we do not perform any de-blending of the \SiIII\ emission in this region.  

The difference seen between BAL and non-BAL continuum slope is due to the larger amount of reddening in BAL, and particularly LoBAL, sources.  This was also seen by Weymann et al. (1991).  Those authors also identified the stronger \FeII\ emission in BAL sources, and suggested that this is also the cause of the apparently stronger \AlIII\ emission.  However, we still see the slightly stronger \AlIII\ here after \FeII\ subtraction, which Weymann et al. (1991) did not do due to lack of a good \FeII\ template at the time.

The similarities between the BAL and non-BAL property distributions can also provide important information.  Three of these are shown in the bottom panels of Figure~\ref{propsfig}; the distributions of $M_{bh}$, $L_{bol}$, and $F_{edd}$ are all visually and statistically similar.  We also do not see stronger emission-line blueshifts in the BAL sample, which was suggested by Richards et al. (2011).

\subsection{Correlations}
Important clues to the structure of BAL quasars could be found by searching for correlations between orientation indicators, outflow parameters, and other properties.  To this end, we have performed an extensive search of such correlations, using the Spearman rank correlation statistic.  The results are presented in Table~\ref{allcorrtbl}; columns 1 and 2 list the parameters being compared, columns 3-5 give the statistics using all (both BAL and non-BAL) sources, including the number ($n$) of sources included, the Spearman rank coefficient ($r_s$), and the associated significance of the coefficients deviation from 0 ($P_{r_s}$).  Columns 6-8 give the results considering only the BAL sample, and columns 9-11 give the results using the non-BAL sources only.  Our two measures of $\alpha_{rad}$ are extremely strongly correlated ($P_{r_s} = 10^{-19}$), and so we only present correlations using $\alpha_{8.4}^{4.9}$ because the results are very similar when using $\alpha_{fit}$.  However, there are a few instances in which the results are quite different (one satisfies our significance cutoff and the other does not), and in these cases we present both results.  Similarly, we investigated correlations with absorption parameters using both our measurements and those in G09, but in no cases did using the G09 values make correlations change from significant to insignificant (or vice-versa) relative to our cutoff, and so we only present the results using our measurements.

\begin{figure}
\centering
 \includegraphics[width=6.5in]{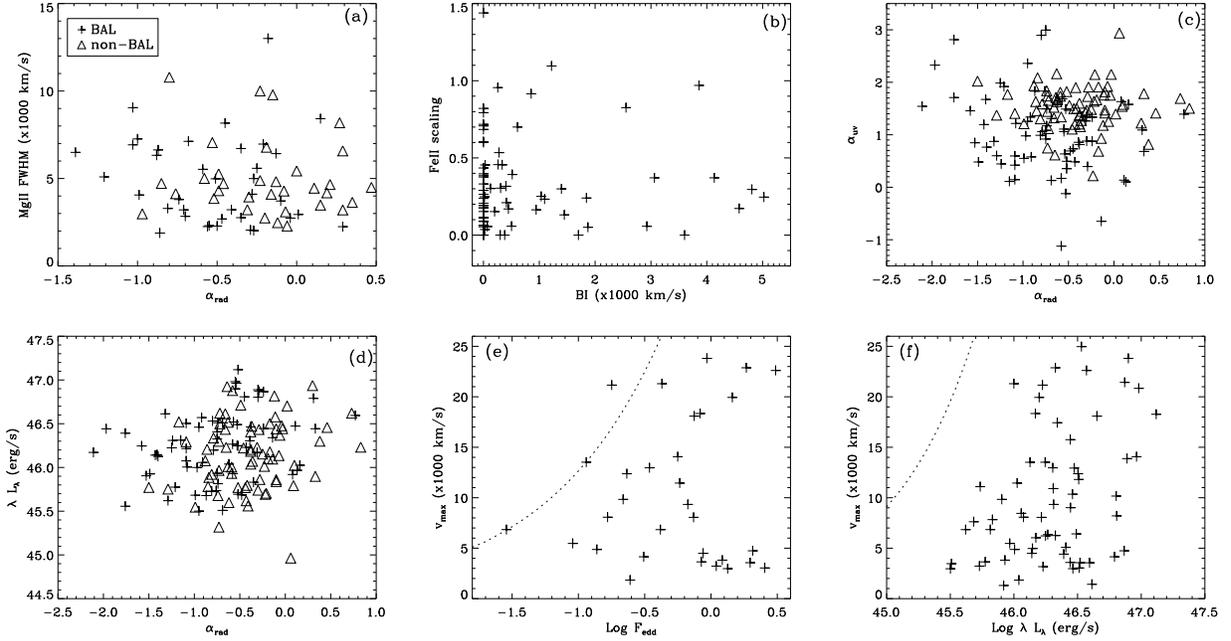}
 \caption{Some of the important correlations (or non-correlations) discussed in the text.  (a) The relationship between $\alpha_{rad}$ ($\alpha_{fit}$ in this case) and \MgII\ FWHM.  While the correlation is only statistically significant at the 1$\sigma$ level, it may simply be due to too few sources as visually it looks quite similar to the similar plot of Fine et al. (2011).  (b) The lack of a correlation between $BI$ and \FeII\ strength, which has been seen by several previous authors in radio-quiet samples.  (c) The lack of a correlation between $\alpha_{uv}$ (a proxy for the amount of dust reddening) and $\alpha_{rad}$.  (d) While marginally significant statistically, visually the correlation between $\alpha_{rad}$ and luminosity is not very striking.  (e) $F_{edd}$ and $v_{max}$ do no correlate, but the relationship is roughly consistent with the upper envelope (dotted line) discussed by Ganguly et al. (2007).  (f) The correlation between luminosity and $v_{max}$, which falls well within the envelope (dotted line) calculated by Ganguly et al. (2007).}
 \label{corrfig}
\end{figure}

Very few significant correlations exist between the properties examined here:  
\begin{enumerate}
\renewcommand{\theenumi}{(\arabic{enumi})}
\item A positive correlation between $\alpha_{8.4}^{4.9}$ and $\lambda L_{2500}$ (and therefore $L_{bol}$ as well) in the BAL sample, though it rises above our significance cutoff when $\alpha_{fit}$ is used.  The correlation is marginally significant for the total and non-BAL samples, but does not satisfy our significance level.  Fine et al. (2011) identified a similar effect in their quasar sample.  This correlation is shown visually in Figure~\ref{corrfig} panel (d), and it does not appear very strong despite the statistics.  If we look at average luminosities of the steep spectrum ($\alpha \le -0.7$ and flat spectrum ($\alpha \ge -0.3$) sources, flat spectrum objects are about 16\% more luminous.

\item An anti-correlation between $\alpha_{rad}$ and $R^{*}$ in the non-BAL sample.  The correlation is marginally significant when considering just $\alpha_{fit}$ in the BAL sample, but becomes insignificant for $\alpha_{8.4}^{4.9}$.  

\item A positive correlation between $v_{max}$ and $\lambda L_{2500}$ (or $L_{bol}$); more luminous sources have higher outflow velocities.  Ganguly et al. (2007) identified a similar relationship, but as an upper envelope.  Our sample only covers the high luminosity end of their sample; see section 5.  This correlation is shown in panel (f) of Figure~\ref{corrfig}, along with the Ganguly et al. (2007) envelope.

\item An anti-correlation between the Eddington fraction and $M_{bh}$.  This is not surprising as SDSS is a flux-limited survey.

\end{enumerate}

Of course, the lack of correlations can provide important clues as well, ruling out certain models.  This will be discussed in section 5.  We point out here a few instances in which our findings do not agree with other work.  We do not see the correlation between $v_{max}$ and $F_{edd}$ found by Ganguly et al. (2007), though we cover a wide range of $F_{edd}$ and outflow velocities.  However, when examining the relationship by eye it is roughly consistent with the envelope they discuss, as shown in panel (e) of Figure~\ref{corrfig}.  We also do not see the strong correlation between \MgII\ FWHM and $\alpha_{rad}$ found by other groups such as Fine et al. (2011) and Jarvis \& McLure (2006), though it is present at about a 1$\sigma$ level if $\alpha_{fit}$ is used.  However, visual inspection of Figure 4a of Fine et al. (2011) and panel (a) of Figure~\ref{corrfig} here show that the relationships look very similar.  It is possible that with more sources here the statistical significance would rise to similar levels.  Weymann et al. (1991) identified a strong correlation between $BI$ and and \FeII\ strength, which we do not see, using either $BI$ or $AI$.  This is shown in Figure~\ref{corrfig} panel (b).  Richards et al. (2011) found that BAL sources with larger \CIII\ and \CIV\ blueshifts also had larger values of $BI$, which they suggests separates BAL sources and radio-loud quasars (which generally have small blushifts) into separate populations.  We see no such trend here in these radio-loud BALs.

\subsection{Composite spectra}
Even if statistical tests produce null results, composite spectra are another way of comparing subsamples of objects to search for important differences.  They can also be used to support the fact that ``significant'' correlations are indeed real.  We present several composite spectra of various subsets of the BAL and non-BAL samples, and all composites are created in the same manner.  

\begin{figure}
\centering
 \includegraphics[width=6.5in]{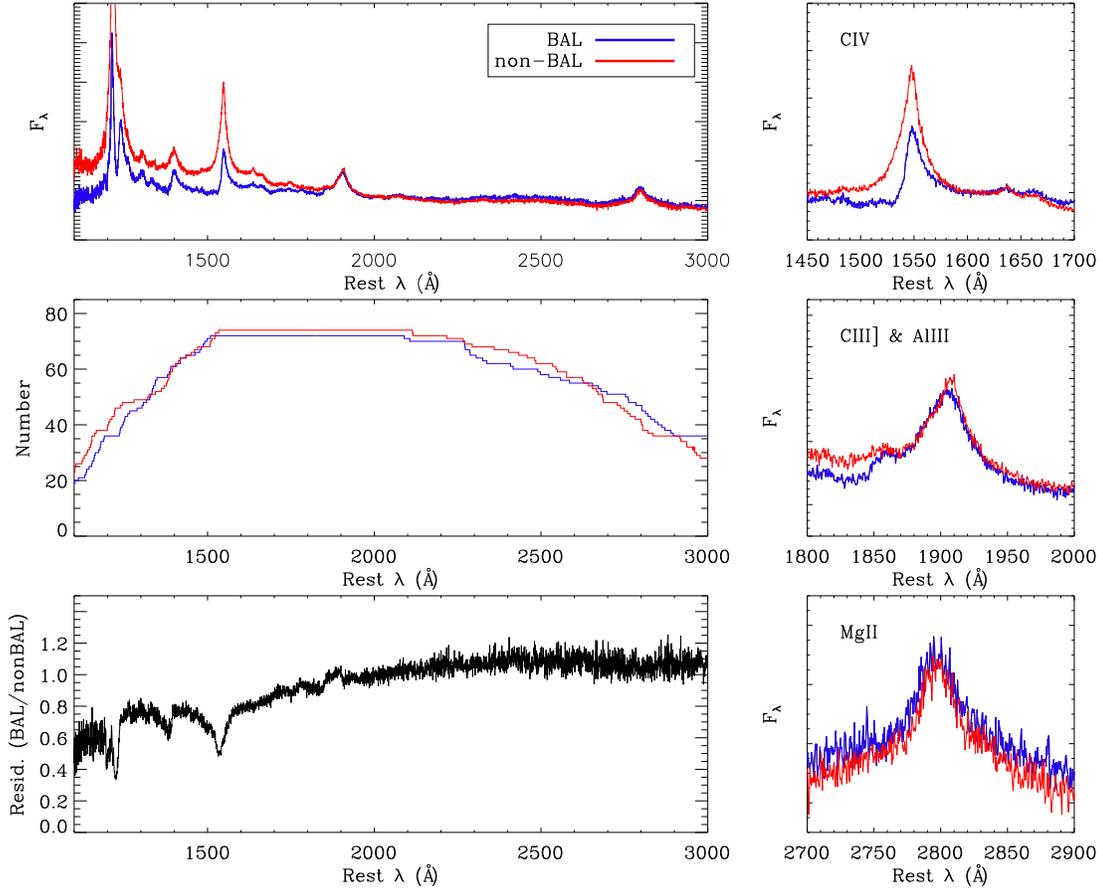}
 \caption{The top left panel shows the composite spectra of the total BAL and non-BAL samples (rescaled to the same flux density from 2000-2050 \AA), the middle left panel shows the number of spectra from each class contributing to the composites at each wavelength, and the bottom left panel shows the residuals of the top spectra.  The panels on the right show close-ups of the \CIV, \CIII/\AlIII, and \MgII\ regions.  In the \CIV\ close-up panel at top right the spectra have been normalized to the same flux density at 1600-1625 \AA\ for easier comparison of the lines.  Fluxes are in arbitrary units.  See the online journal for the color versions of this and the following composite spectra.}
 \label{balnbalcomp}
\end{figure}

\begin{figure}
\centering
 \includegraphics[width=6.5in]{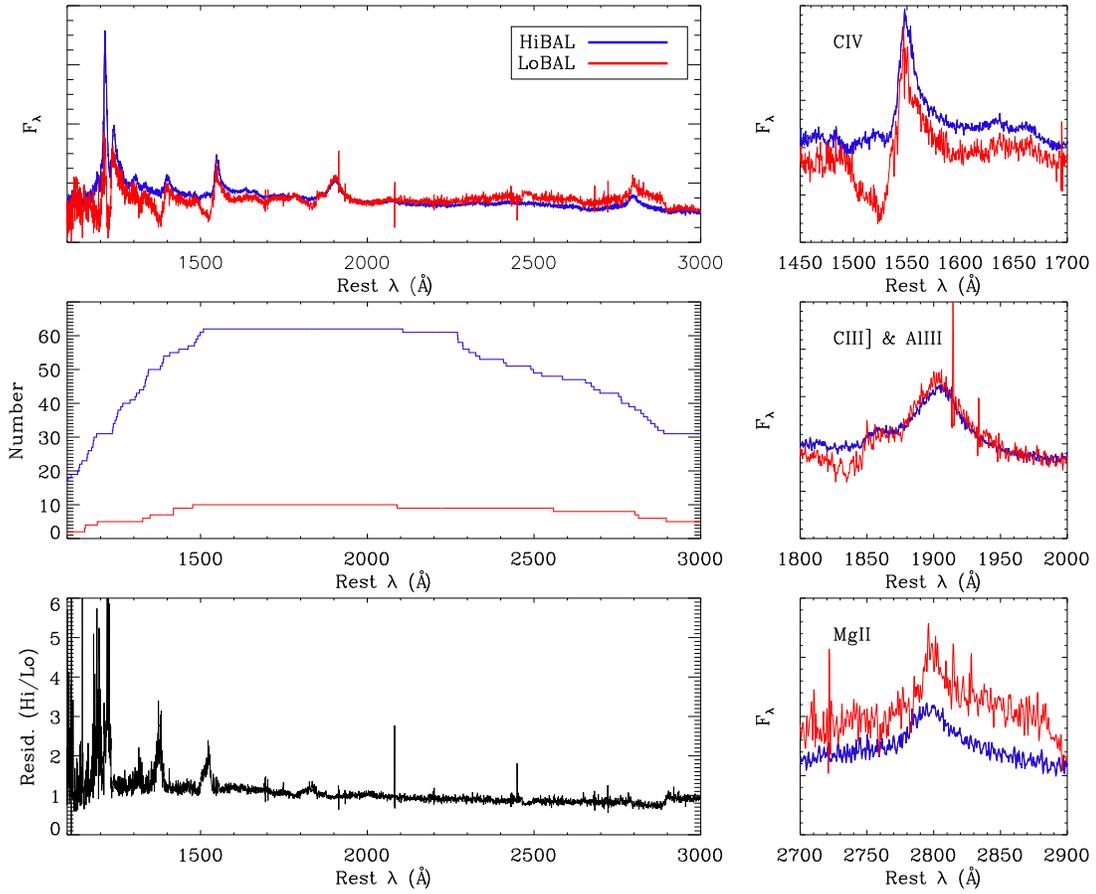}
 \caption{The top left panel shows the composite spectra of the HiBAL and LoBAL subsamples; other panels are the same as Figure~\ref{balnbalcomp}.}
  \label{hilocomp}
\end{figure}

\begin{figure}
\centering
 \includegraphics[width=6.5in]{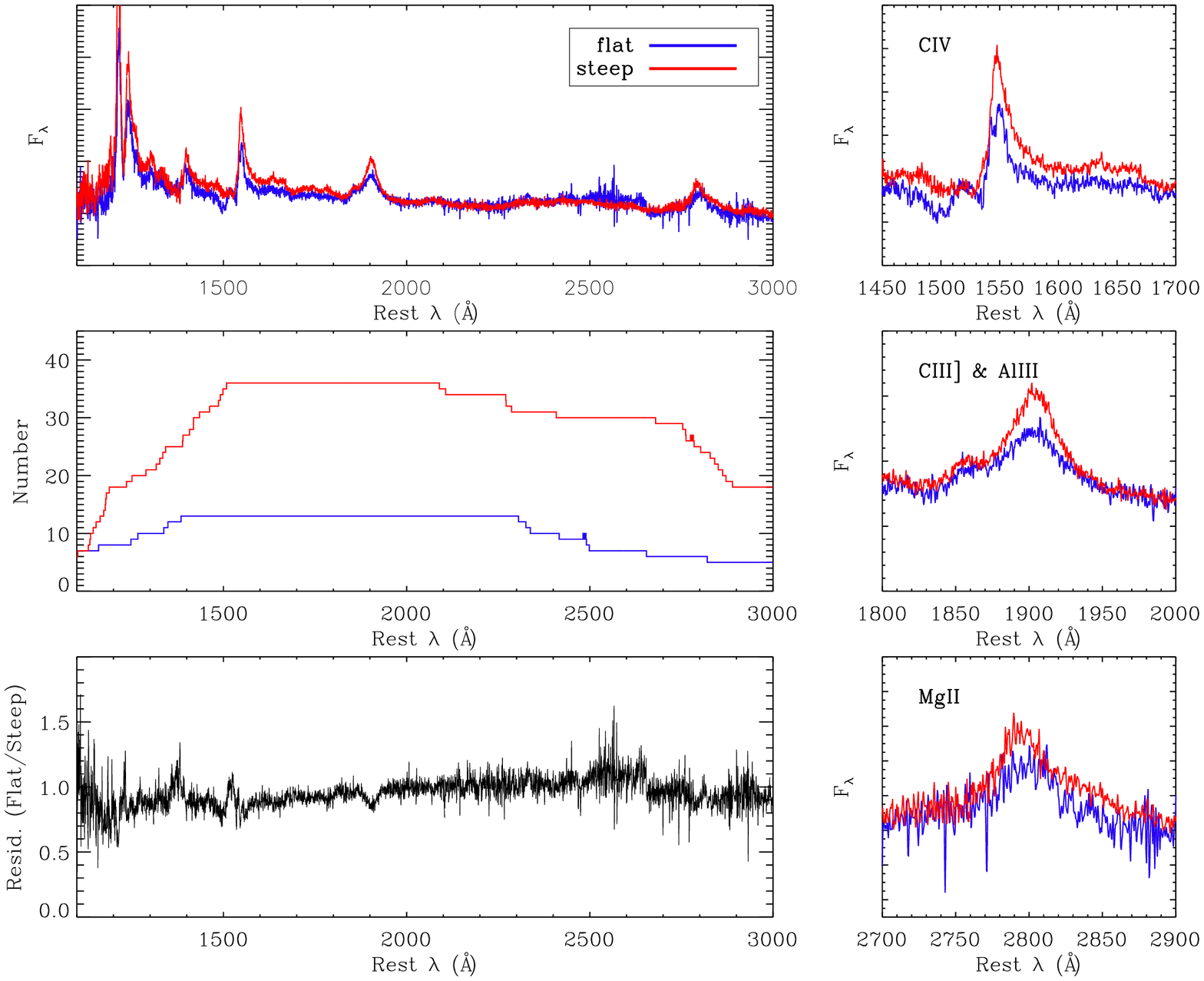}
 \caption{The top left panel shows the composite spectra of steep ($\alpha_{8.4}^{4.9} < -0.7$) and flat ($\alpha_{8.4}^{4.9} > -0.3$) spectrum BAL sources; other panels are the same as Figure~\ref{balnbalcomp}.} 
 \label{balalphacomp}
\end{figure}

\begin{figure}
\centering
 \includegraphics[width=6.5in]{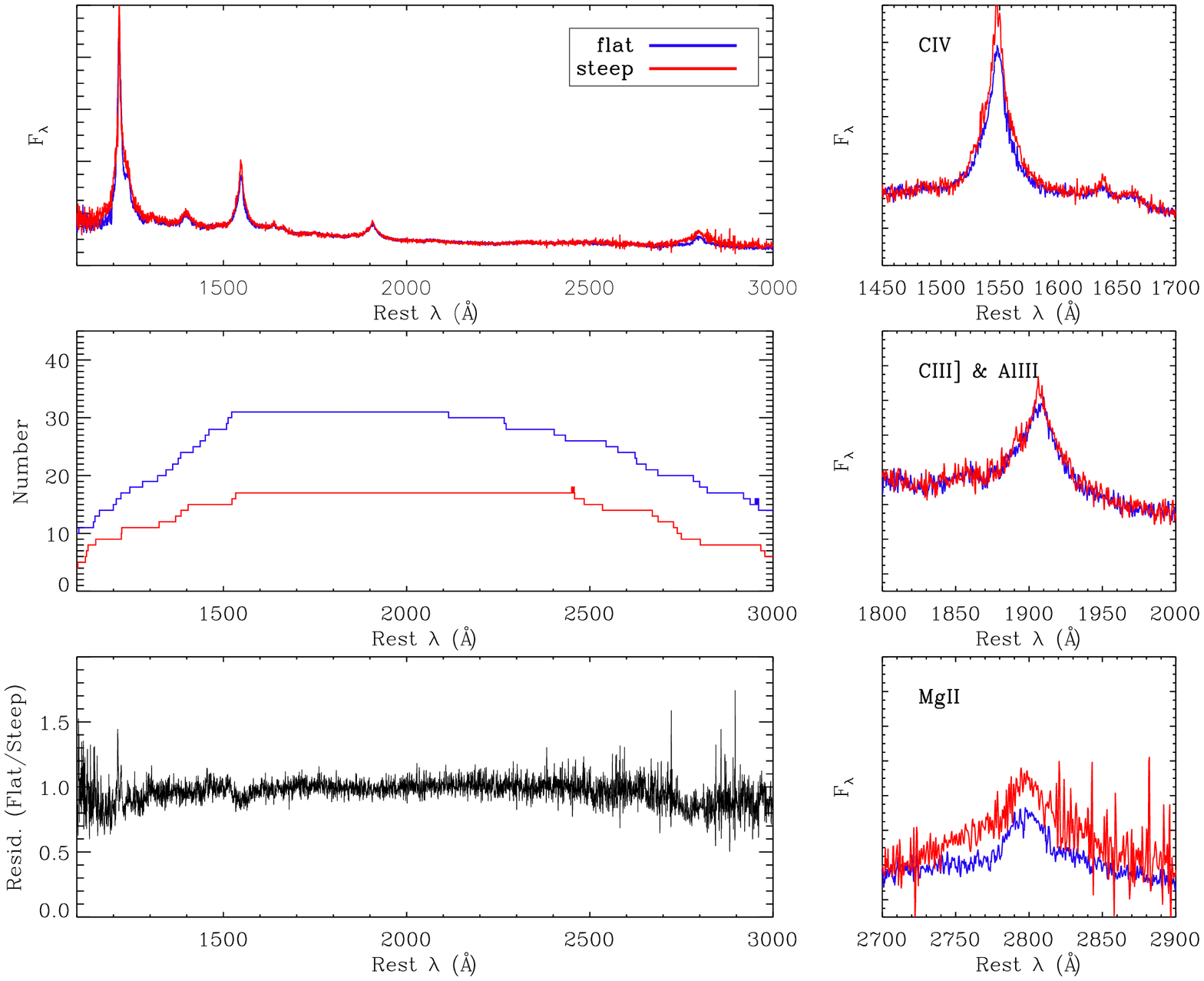}
 \caption{The top left panel shows the composite spectra of steep ($\alpha_{8.4}^{4.9} < -0.7$) and flat ($\alpha_{8.4}^{4.9} > -0.3$) spectrum non-BAL sources; other panels are the same as Figure~\ref{balnbalcomp}.} 
 \label{nbalalphacomp}
\end{figure}

\begin{figure}
\centering
 \includegraphics[width=6.5in]{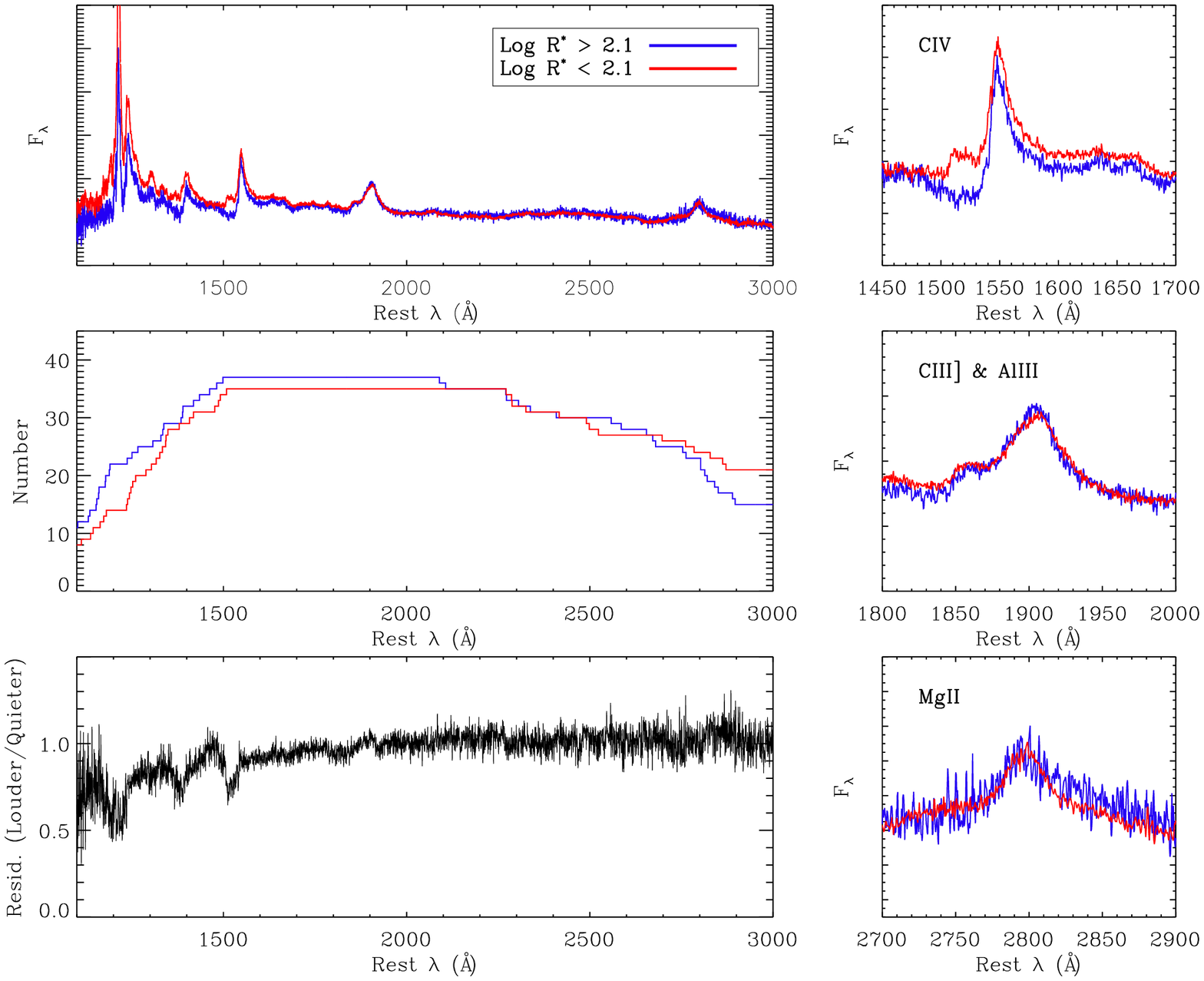}
 \caption{The top left panel shows the composite spectra of the radio-loudest ($\log R^{*} > 2.1$) and radio-quietest ($\log R^{*} < 2.1$) BAL sources; other panels are the same as Figure~\ref{balnbalcomp}.} 
 \label{balloudcomp}
\end{figure}

\begin{figure}
\centering
 \includegraphics[width=6.5in]{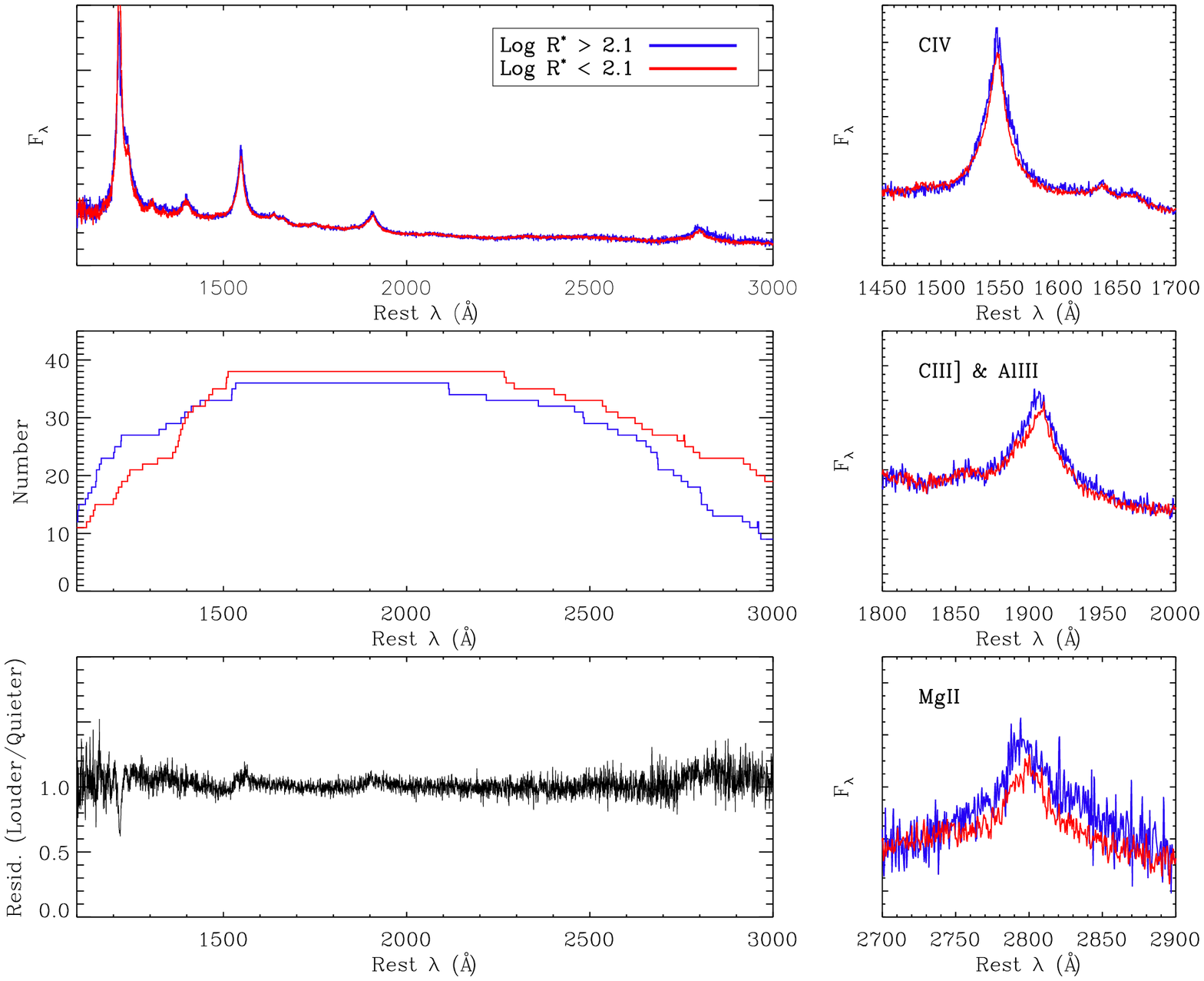}
 \caption{The top left panel shows the composite spectra of the radio-loudest ($\log R^{*} > 2.1$) and radio-quietest ($\log R^{*} < 2.1$) non-BAL sources; other panels are the same as Figure~\ref{balnbalcomp}.} 
 \label{nballoudcomp}
\end{figure}

Spectra contributing to the given composite are rescaled to have the same mean flux density in the wavelength region 2000-2050\AA.  This region was selected because it is one of the few continuum regions in all of the spectra.  The average of all of the spectra are then taken (medians were also checked, but they are not presented because they are similar to the average), using a 3$\sigma$ clipping.  The composites compared are:
\begin{enumerate}
\renewcommand{\theenumi}{(\arabic{enumi})}
 \item BAL versus non-BAL, shown in Figure~\ref{balnbalcomp}.  From these plots we can see that BAL quasars are generally more reddened, have weaker \CIV, \SiIV, and likely \NV.  
 The \AlIII\ emission does not appear significantly stronger in the BAL sources, contrary to the KS/RS tests on the EW distributions.  We can also see that the seemingly stronger/wider \CIII\ emission is indeed due to stronger \SiIII\ in the BAL sources, as the peak of the blend is at slightly bluer wavelengths.  The stronger \FeII\ emission between \CIII\ and \MgII\ in the BAL composite is also visible.  If the emission lines are normalized to the same heights, no new details are apparent (lines have indistinguishable widths, etc.).
 
 \item HiBAL versus LoBAL, shown in Figure~\ref{hilocomp}.  Detailed comparisons are difficult here, as there are only ten LoBALs in the sample, and at some wavelengths not all of them contribute.  However a few things are apparent.  LoBALs are indeed more red, but emission lines appear to be of similar strengths.  The deep BAL troughs remain in the LoBAL composite, but largely cancel out in the one for HiBALs.  This indicates a more consistent BAL profile in LoBALs, as opposed to the varying types of BALs seen in HiBALs, but again this could be due to low numbers.  Our results confirm those of Brotherton et al. (2001).
 
 \item Flat ($\alpha_{8.4}^{4.9} \geq -0.3$) versus steep ($\alpha_{8.4}^{4.9} \leq -0.7$) radio spectrum, shown in Figure~\ref{balalphacomp} (BALs) and Figure~\ref{nbalalphacomp} (non-BALs).  We did a similar composite using $\alpha_{fit}$, but it is not shown here as the results were very similar.  In the BAL comparison, though the number of flat spectrum sources is small, it appears that flat spectrum sources may actually be reddened slightly compared to steep spectrum sources.  Note that this is not due to LoBAL contributions (because they are more reddened), as 5 LoBALs are included in the steep spectrum composite, while only 2 contribute to the flat spectrum.  Therefore, leaving out LoBALs actually enhances this effect.  The flat sources have weaker emission lines, in particular \CIV\ and \CIII.  The flat spectrum sources also tend to appear more luminous, likely due to the accretion disk orientation (see section 5) and not intrinsic luminosity differences, which would rule out the weaker emission lines being due to the Baldwin effect.  There is also some indication that the flat spectrum sources have both higher maximum outflow velocities and lower minimum velocities, but again we caution this is very possibly due to the number of sources.  If the lines are normalized, the \MgII\ line appears slightly wider in the steep spectrum sources.  In the non-BAL flat versus steep comparison, we see the two composites are much more similar.  \CIV\ is a small amount stronger in the steep spectrum sources, and \MgII\ is slightly wider.
 
 \item ``Radio-loudest'' ($\log R^{*} > 2.1$) versus ``radio-quietest'' ($\log R^{*} < 2.1$), shown in Figure~\ref{balloudcomp} (BALs) and Figure~\ref{nballoudcomp} (non-BALs).  The division at $\log R^{*}=2.1$ was simply chosen because it divides the sources into roughly equal groups.  In the BAL composites, we see that the radio-loudest sources are actually slightly more red than the radio-quieter ones, but their emission lines are essentially the same.  It also appears that the radio-loudest sources have less detached, more consistent BAL troughs.  The radio-quieter sources also have low velocity outflows, as you can see by the shape of the emission line, but BAL profiles must vary more in velocity in order to average out of the composite.  In the non-BAL composites, once again the spectra are much more similar.  The higher reddening in the radio-loudest sources is not seen,  but there is a slight enhancement in emission-line strengths in these objects.
\end{enumerate}

Please see the online version of the journal for the color versions of these composite spectra.

\section{DISCUSSION}
The key results of this work are the lack of strong differences between radio-loud BAL and non-BAL properties, and the lack of correlations between $\alpha_{rad}$ (orientation) and outflow properties.  It is often questioned whether or not we can treat radio-loud and radio-quiet BAL quasars in the same way, or if they form distinctly separate populations.  While we need a detailed comparison of radio-loud and radio-quiet BAL quasars to do this properly, our results here are very similar to those for the comparisons between radio-quiet BALs and non-BALs.  Overall, the properties are the same with the exception of more reddening in BAL sources, stronger \FeII\ emission, and likely stronger \SiIII\ (manifested in the differences seen in \CIII).  It is also clear from the composite spectra (Figure~\ref{balnbalcomp}) that \CIV\ emission is weaker in BAL quasars, which is difficult to explain via the Baldwin effect because the samples are well matched in luminosity.  However, it could be related to the tendency for BAL quasars to lie at one end of eigenvector 1.  The fact that these results are similar to comparisons between radio-quiet objects indicates that the ultraviolet properties of radio-loud and radio-quiet BAL quasars are the same.  This is supported by other work as well, such as the similarity in optical polarization properties in radio-loud and radio-quiet BAL quasars (DiPompeo et al. 2010).

The significant correlations we found are overall not very telling.  The correlation between $\alpha_{rad}$ and $\lambda L_{2500}$ ($L_{bol}$) shows that more pole-on sources tend to have higher luminosities, which could just be due to the orientation of the accretion disk.  As stated above, flat spectrum sources are on average about 15\% more luminous.  If we assume flat spectrum sources are seen on average around 10 degrees from the accretion disk axis, and steep spectrum sources are seen around 30 degrees (see for example DiPompeo et al. 2012), then this luminosity difference is consistent with an effect due to accretion disk viewing angle (see Nemmen \& Brotherton 2010).  This correlation is only significant for one measure of $\alpha_{rad}$ and in the BAL sample, though a similar finding was made by Fine et al. (2011).  It is somewhat unexpected to see the anti-correlation between $\alpha_{rad}$ and $R^{*}$, where the more edge-on sources are more radio-loud.  One would expect the opposite to occur, as beaming should cause the enhancement of radio emission to be much stronger than any enhancement of ultraviolet emission in the face-on sources and thus raise $R^{*}$.  It is also unclear why this correlation is so much stronger in the non-BAL sample.

There appear to be more interesting results in the non-correlations than the correlations themselves, and these non-correlations can provide some constraints on models suggested in other works.  For example, there is no strong correlation between $\alpha_{rad}$ and $\alpha_{uv}$ (a reddening indicator), as shown in panel (c) of Figure~\ref{corrfig}.  This is again consistent for what was found by Fine et al. (2011).  In the traditional dusty ``torus'' model, we might expect the steeper radio spectrum sources to show more reddening.  Despite the lack of a correlation, it appears that in the flat/steep BAL composite comparisons in Figure~\ref{balalphacomp} flatter $\alpha_{rad}$ sources are slightly redder, and this is clearly not present in the flat/steep non-BAL composite of Figure~\ref{nbalalphacomp}.  The lack of this effect in the non-BAL sample could support a picture in which a dusty torus is still present, but no viewing angles to unabsorbed quasars are near enough to it to see any significant reddening effect on the spectrum.  This fits with the results of DiPompeo et al. (2012), where at the largest viewing angles one will only see BAL sources, and no unabsorbed sources.  

However, it is interesting that in the BAL sample the composites suggest that flatter spectrum sources are on average more reddened.  It could simply be that there are not enough flat spectrum BAL sources (only 13 with $\alpha_{8.4}^{4.9} > 0.3$), and this effect seen in the composites is not actually real.  On the other hand, if the effect is real, it could have implications for the amount of dust along different lines of sight and possible covering fractions.  These observations could support the model suggested by DiPompeo et al. (2012), where orientation and evolution both play a role; BAL quasars begin completely enshrouded and absorption can be seen along all lines of sight, but as they evolve polar regions are cleared of material first in the area around the developing radio jet.  Thus when polar BALs are seen, they are seen in younger, fully enshrouded systems that are more reddened.  While Gallagher et al. (2007) found that on average BAL quasars do not seem to have a significant IR excess, ruling out that all BAL quasars are completely enshrouded, it would be interesting to analyze this as a function of spectral index (viewing angle) to see if the flat spectrum BALs show an IR excess compared to steep spectrum BALs.  Additionally, the results of Allen et al. (2011) indicate that the redshift range around 2-3 is when the BAL phenomenon peaks, and therefore this is the time when most BAL winds ``turn on''.  Thus, in the above picture, we would expect most polar BALs to be seen in this redshift range.  Unfortunately, the number of flat spectrum BALs in this sample is too small to really verify this.  However, the redshift range of the possibly polar BALs found by short timescale radio variability (Ghosh \& Punsly 2007, Zhou et al. 2006) does appear to peak at around $z=2.5$.

Another interesting thing to notice in the flat/steep composites, is that the steep spectrum (more edge-on) sources seem to have stronger emission lines.  This is seen more so in the BAL sample, but it is present in the non-BALs as well, most clearly for \CIV.  This difference could again be due to the relative numbers contributing to the composites.  Fine et al. (2011) saw this effect in the narrow emission lines of \OII\ and \OIII.  They interpreted this as having spherical NLR clouds emitting mostly from the cloud face closest to the accretion disk, with cooler, more optically thick upper halves (see their Figure 11).  Our results here suggest that something similar may be happening in the BLR clouds as well.  

All of these sources are high mass, with an average mass over $10^{9}\ M_{\odot}$.  The fact that they all have such high masses supports the idea that black hole mass does play into the ability of a quasar to become a powerful radio source.  However, there is no correlation between $M_{bh}$ and $R^{*}$ or $L_r$ in either sample, suggesting that once a source is massive enough to produce a radio jet, other factors become more important in determining its power.  This could for example be black hole spin (e.g. Richards et al. 2011).  Ganguly et al. (2008) concluded that BAL sources were not likely super-accretors, and we confirm that here for radio-loud sources.  However, despite the statistical results the highest Eddington fraction sources are in the BAL sample.

The luminosities probed are also quite high, and our results suggest that radiation pressure is a significant driver of the winds in these sources, manifested in the correlation between $v_{max}$ and luminosity (panel (f) of Figure~\ref{corrfig}).  The envelope in this relationship presented by Ganguly et al. (2007) really requires lower luminosities to see, however our results are clearly consistent with their findings.  Also, the fact that no outflow properties seem to correlate with orientation indicates that the outflows in these sources have the same driving force.  In the scenario presented by Richards et al. (2011), where radiation pressure dominated sources tend to have ``equatorial'' winds (moving away from the jet axis), and other sources may have winds dominated by some other mechanism (such as magneto-hydrodynamic forces), we might expect to identify some trends with viewing angle.  For example, in this picture we expect to see $BI$ and $\alpha_{rad}$ to correlate, in the sense that the highest $BI $ sources should be seen at larger inclination angles.  This is not the case.  Additionally, we do not see the larger blueshifts in BAL emission lines seen by Reichard et al. (2003), though we do see the decrease in \CIV\ EW.  This could be because our sources are all radio-loud.  In the scenario presented by Richards et al. (2011), sources with large blueshifts are radiation pressure dominated, X-ray weak, and radio quiet.  However, placing these radio-loud BAL quasars into that type of picture appears to be difficult.

\section{SUMMARY}
We have performed fits of the SDSS DR7 spectra and an analysis of the ultraviolet properties (and the subsequently derived intrinsic properties) of the 74 radio-loud BAL and 74 radio-loud non-BAL quasars presented in the spectral index analysis of DiPompeo et al. (2011).  Measurements of both samples include emission-line EW, FWHM, and blueshift of \CIV\ (non-BAL only), \AlIII, \CIII, and \MgII, \FeII\ strength, and continuum shape.  Additionally, we measure $BI$, $AI$, $v_{min}$, and $v_{max}$ of the \CIV\ BAL in the BAL quasar sample.  From the fits, we calculate the bolometric luminosity, black-hole masses (when \MgII\ is in the spectral window), and Eddington fractions.  We employed two statistical tests on the distributions of these properties to look for differences between radio-loud BAL and non-BAL quasars.  We searched for correlations between numerous parameters, most importantly between radio spectral index (as a proxy for viewing angle) and absorption properties.  We also made comparisons using composite spectra of various subsets of each sample.

We summarize the main results as follows:
\begin{enumerate}
\renewcommand{\theenumi}{(\arabic{enumi})}
\item Radio-loud BAL and non-BAL quasars have similar ultraviolet and intrinsic properties, and the (generally small) differences seen are the same as when comparing radio-quiet objects.  Radio-loud BALs have redder spectra, weaker \CIV\ emission, and stronger \FeII\ emission compared to radio-loud non-BALs.  

\item No significant correlations exist between radio spectral index and BAL trough properties.  It is now fairly well established that BAL outflows can occur along any line of sight, and this suggests that all of them have similar properties and thus are driven by similar mechanisms.

\item Composite spectra indicate that BAL sources with flat radio spectra appear more reddened.  This is counter to what we would expect in the traditional dusty torus models, and indicates that BAL quasars with more polar outflows have a higher dust content.  Non-BAL quasars on the other hand have no relationship between spectral index and reddening, which indicates that they are not generally seen from any viewing angle that results in significant reddening.

\item There is a significant correlation between ultraviolet (and bolometric) luminosity and maximum outflow velocity.  The relationship is also consistent with the upper envelope identified by Ganguly et al. (2007).  There is no preference for flat or steep $\alpha_{rad}$ sources to fall in particular places along this relationship.  This suggests that radiation pressure is the dominant force in driving BAL outflows at any viewing angle.

\item Radio-loud BAL quasars are generally high black hole mass and high luminosity, but have fairly typical Eddington fractions.  However, the non-BAL sources selected as a matched sample have these properties as well, and so there is no significant difference between the two samples.
\end{enumerate}

These and other recent results show that BAL quasars cannot be explained using large viewing angle or evolutionary explanations only.  A more complex picture is needed, perhaps one that incorporates both of these ideas.

\section*{acknowledgments}
We thank the anonymous referee for useful comments that significantly improved this paper.  We also thank Carlos De Breuck for his work in putting this sample together and performing the radio observations.  We also thank Rajib Ganguly and Zhaohui Shang for providing files and code for the fitting procedures.

\clearpage

\clearpage

\setcounter{table}{0}
\begin{table}
 \tiny
  \caption{Sample properties.}
  \label{proptbl}
  \begin{tabular}{lcccccccccc}
  \hline
  Object & $z$  & $\log R^{*}$ & $\log L_{r}$ & $\log L_{bol}$ & $\log M_{bh}$  & $F_{edd}$ & $BI$ & $AI$ & $v_{max}$ & $v_{min}$  \\
   &      &     &  (erg/s/Hz)    &  (erg/s) & (M$_{\odot}$)  &  & (km/s) & (km/s) & (km/s) & (km/s)   \\
   (1) & (2) &   (3) & (4) &   (5) & (6) &   (7)  & (8)  &  (9)   &  (10)  &  (11)  \\
 \hline
SDSS J001408.22-085242.2 & 1.74 & 2.27 & 33.1 & 46.6 &  9.6 & 0.09 &    402 &   2910 &   5469 &    1355 \\
SDSS J002440.99+004557.7 & 2.24 & 3.04 & 33.8 & 46.2 & ... & ... &      0 &    216$^1$ &   2960 &    1349 \\
SDSS J003923.18-001452.6 & 2.23 & 2.51 & 33.4 & 46.4 &  8.3 & 1.04 &      0 &    104 &   3225 &    1138 \\
SDSS J004444.06+001303.5 & 2.29 & 3.06 & 33.8 & 46.4 & ... & ... &     25 &    146 &   7620 &    4726 \\
SDSS J014847.61-081936.3 & 1.68 & 2.58 & 33.5 & 46.6 &  9.2 & 0.21 &   1396 &   2736 &   9836 &    1197 \\
SDSS J024534.07+010813.7 & 1.54 & 3.25 & 34.1 & 46.6 &  9.6 & 0.07 &      0 &   $>$127 &  $>$2397 &      -7 \\
\hline
SDSS J000050.60-102155.9 & 2.64 & 1.65 & 33.5 & 47.2 & ... & ... &    ... &    ... &    ... &     ... \\
SDSS J000221.11+002149.3 & 3.07 & 1.75 & 33.4 & 47.1 & ... & ... &    ... &    ... &    ... &     ... \\
SDSS J001507.00-000800.9 & 1.70 & 1.54 & 32.9 & 47.0 &  8.8 & 1.08 &    ... &    ... &    ... &     ... \\
SDSS J073659.31+293938.4 & 2.45 & 2.02 & 33.1 & 46.4 & ... & ... &    ... &    ... &    ... &     ... \\
\hline \\
   \end{tabular}
    \\
{
\raggedright
 $^1$$AI$ measured using a criteria of continuous absorption for 800 km/s, instead of 2000 km/s.

Only a portion of the table is shown here as an example; the full table will be published with the online version of MNRAS.  The full names of the objects in the sample and their SDSS redshifts are given in the first two columns.  Columns (3)-(7) give some calculated properties; see Sections 3.1-3.3 for detailed discussion of these values.  The final four columns give measurements of the \CIV\ BAL trough; see Section 3.4 for details.  In some cases the regions of possible absorption are redshifted out of the spectrum (for $z < 1.6$), and so lower limits are given.The BAL and non-BAL samples are separated by the horizontal line. \\
   }
\end{table}

\begin{landscape}
\setcounter{table}{1}
\begin{table}
 \tiny
  \caption{Line measurements.}
  \label{lineproptbl}
  \begin{tabular}{lcccccccccccccc}
  \hline
  Object & $f_{1000}$$^{1}$ & $\alpha_{uv}$ & \FeII $^{2}$ & \CIV\ EW & \CIV\ FWHM & \CIV\ BS$^3$ & \AlIII\ EW & \AlIII\ FWHM  & \CIII\ EW & \CIII\ FWHM  & \CIII\ BS$^3$ & \MgII\ EW & \MgII\ FWHM & \MgII\ BS$^3$  \\
   &      &     &      &  (\AA) & (km/s) & (km/s)  & (\AA) & (km/s) & (\AA) & (km/s) & (km/s)  & (\AA) & (km/s) & (km/s) \\
      (1) & (2) &   (3) & (4) &   (5) & (6) & (7)  &  (8) & (9) &   (10) & (11) &   (12) & (13)  & (14)  &  (15) \\
 \hline
0014-0852 &   7.0 &  0.10 & 0.32 & ... &   ... &  ... &   1.2 $\pm$   0.3 &  2523 $\pm$   34 &  13.2 $\pm$   0.7 &  4101 $\pm$  103 &  752.9 & 42.4 $\pm$   4.7 &  8419 $\pm$  485 &  -91.7\\
0024+0045 &   3.1 &  1.25 & 0.00 & ... &   ... &  ... &   0.0 &   ... &  32.1 $\pm$   5.9 &  2686 $\pm$  508 &  418.3 &... &   ... &  ...\\
0039-0014 &   4.2 &  0.99 & 0.11 & ... &   ... &  ... &   0.0 &   ... &  27.1 $\pm$   3.1 &  3030 $\pm$  281 &  515.3 & 37.3 $\pm$  11.8 &  2055 $\pm$  507 &   27.5\\
0044+0013 &   2.7 &  0.69 & 0.04 & ... &   ... &  ... &   5.0 $\pm$   4.4 &  5859 $\pm$  281 &  17.6 $\pm$   2.1 &  6873 $\pm$  691 &  564.2 &... &   ... &  ...\\
\hline
0000-1021 &  26.8 &  1.65 & 0.00 &  58.4 $\pm$   1.5 &  3054 $\pm$   78 &  306.8 &   4.5 $\pm$   0.7 &  9934 $\pm$  103 &  14.9 $\pm$   0.6 &  4102 $\pm$   79 &  329.3 &... &   ... &  ...\\
0002+0021 &  23.1 &  2.15 & 0.00 &   9.9 $\pm$   0.3 &  3941 $\pm$  154 & 1233.6 &  12.8 $\pm$   3.2 & 10134 $\pm$  155 &  19.0 $\pm$   1.0 &  5012 $\pm$  146 &  -81.0 &... &   ... &  ...\\
0015-0008 &  55.6 &  1.44 & 0.66 &  69.8 $\pm$   1.4 &  4577 $\pm$   72 &  489.6 &   2.4 $\pm$   0.2 &  3556 $\pm$   24 &  17.9 $\pm$   0.5 &  5462 $\pm$   65 &  184.3 & 31.0 $\pm$   1.9 &  2743 $\pm$   69 &  181.6\\
0736+2939 &   4.7 &  1.37 & 0.00 &  48.7 $\pm$   3.5 &  2748 $\pm$  770 &  252.7 &   0.0 &   ... &  23.8 $\pm$   4.7 &  2070 $\pm$  435 &  265.3 &... &   ... &  ...\\
\hline \\
   \end{tabular}
    \\
    {
    \raggedright
    $^{1}$In units of $10^{-17}$ ergs/s/cm$^2$ \AA. \\
   $^{2}$These values are scalings of the \FeII\ flux in the I Zw 1 template. \\
   $^{3}$BS is an abbreviation of BS, in order for the table to fit on the page.

   Only a portion of the table is shown here as an example; the full table will be published with the online version of MNRAS.  Errors on $f1000$, $\alpha_{uv}$, and the \FeII\ scaling are omitted because they are typically quite small, on the order of one to a few percent.  Line parameters are given as ``0'' when the final fit is indistinguishable from a line not being present.  In some cases lines could not be measured, due to various factors like redshift, noisy or missing data, intervening absorption, or BAL features.  See section 3.2 for full details.  The top half of the table is the BAL sample, and the lower half below the horizontal line is the non-BAL sample. \\
  }
\end{table}
\end{landscape}

\clearpage

\setcounter{table}{2}
\begin{table}
 \small
  \caption{BAL and non-BAL property statistics and comparisons.}
  \label{balnbalcomptbl}
  \begin{tabular}{lccccccccccccc}
  \hline
                     & &  \multicolumn{3}{c}{BAL}  & &  \multicolumn{3}{c}{non-BAL} &    &            &                    &         &                   \\          
                        \cline{3-5}                                   \cline{7-9}
 Parameter         &  & $\mu$ & $\sigma$ & $n$ &  & $\mu$ & $\sigma$ & $n$ &    &  $D$  &        $P_{ks}$             &  $Z$ &  $P_{rs}$                    \\
\hline
$\alpha_{uv}$  &  &  1.04    &   0.74       &  72   &  &  1.53   &    0.40       & 74   &    &  0.47  & \textbf{${10^{-8}}$} & 5.23  & \textbf{${10^{-8}}$} \\ 
\FeII                     &  &  0.33    &   0.30       &  72   &  &  0.18   &    0.20       & 74   &    &  0.35  & \textbf{${10^{-4}}$} & -3.46 & \textbf{${10^{-4}}$} \\ 
\CIV\ EW             &  &     ...      &     ...          &  ...    &  &   45.1  &    28.6       & 74   &    &    ...     &               ...                   &     ...   &                ...                  \\
\CIV\ FWHM       &  &     ...      &     ...          &  ...    &  &  4322  &    1488      & 74   &    &    ...     &               ...                   &     ...   &                ...                  \\  
\CIV\ blueshift    &  &    ...       &     ...          &  ...    &  &   395   &      420      &  74   &    &    ...     &               ...                   &     ...   &                ...                  \\  
\AlIII\ EW             &  &    4.36  &    5.89      & 69   &  &   2.78   &    4.60       & 72   &    &   0.28 &           \textbf{0.005}                & 2.59  &          \textbf{0.005}                 \\  
\AlIII\ FWHM       &  & 4912     &   2809      & 55   &  &  6417  &    3899      & 45   &    &  0.22  &           0.143                &  1.95 &          0.025                 \\  
\CIII\ EW             &  &    21.2   &    10.8      & 71   &  &   17.4   &    7.19       & 72   &    &  0.26  &           0.014                & 2.34  &   \textbf{0.009}           \\  
\CIII\ FWHM       &  & 5780     &   1894     & 70   &  &   4851  &    1688      & 72   &    &  0.26  &           0.015                & -3.00 &          \textbf{0.001}                 \\  
\CIII\ blueshift    &  &  579      &   665       & 70    &  &   166    &    695        & 72   &    &  0.21  &           0.078                & -2.01  &   0.022                      \\
\MgII\ EW           &  &   41.2     &   28.1      & 38   &  &   47.2   &    30.9       & 33   &    &  0.29   &           0.087                &-1.20  &          0.115                 \\  
\MgII\ FWHM     &  & 4807      &   2509    & 28   &  &   4978  &    2101      & 33   &    &  0.25  &           0.183                & -0.70 &          0.241                  \\  
\MgII\ blueshift  &  &  24.6      &  266        & 38   &  &  10.8    &    363        & 33   &    &  0.13  &           0.921                & 0.00  &           0.500                 \\
$\log R^{*}$      &  & 2.15       & 0.49        & 72   &  &  2.09     &   0.44        & 74   &   &  0.10  &            0.820               &  -0.66 &          0.254                 \\
$\log L_r$ (erg/s/Hz)        &  &  33.4      &  0.39       & 72   &   &  33.5    &    0.43       &  74  &    &  0.10  &           0.799               &  0.80   &         0.211                  \\ 
$\log L_{bol}$ (erg/s)  &   & 46.92    & 0.37        & 72   &  &  46.80  &   0.37        & 74    &   &  0.16  &           0.262                & -1.69  &         0.045                   \\
$\log M_{bh}$ (M$_{\odot}$)  &   & 9.12      & 0.44        & 38   &   & 9.15    &   0.36         & 33    &   & 0.19   &           0.471               & 0.29   &         0.388                     \\
$F_{edd}$        &    & 0.73     &  0.75       & 38   &    & 0.40   &    0.31        & 33   &    & 0.27  &           0.128                & -1.41  &        0.078                      \\ 
\hline
   \end{tabular}
   \\  
{
\raggedright    
   All EW measures are in units of \AA, and FWHM measures are in km/s.  The mean ($\mu$), standard deviation ($\sigma$), and number ($n$) of BAL or non-BAL sources with a given measurement are presented.  The final four columns show results of statistical tests comparing the BAL and non-BAL distributions, using both K-S and R-S tests.  Results do not change significantly when making a cut of SNR $\geq$ 6, nor when excluding LoBALs from the BAL sample. \\
 }
\end{table}

\clearpage

\begin{landscape}

\setcounter{table}{3}
\begin{table}
\scriptsize
  \caption{Correlation tests.}
  \label{allcorrtbl}
  \begin{tabular}{llccccccccccc}
  \hline
                                       &                             & \multicolumn{3}{c}{All}                                           &   &  \multicolumn{3}{c}{BAL}                                          &   &   \multicolumn{3}{c}{non-BAL}  \\
                                                                           \cline{3-5}                                                                      \cline{7-9}                                                                          \cline{11-13}
Param 1                        &       Param 2      &  $n$ & $r_{s}$ & $P_{r_s}$ &   & $n$ & $r_{s}$ & $P_{r_s}$ &   & $n$ & $r_{s}$ & $P_{r_s}$   \\
\hline
$\alpha_{8.4}^{4.9}$  & $\alpha_{fit}$   &   144 &  0.66    & \textbf{${10^{-19}}$} &   & 72    &   0.65   & \textbf{${10^{-10}}$} &   & 72   & 0.64  & \textbf{${10^{-9}}$}    \\[2pt]
$\alpha_{8.4}^{4.9}$  &  \CIV\ FWHM    &    ...    &     ...      &       ...          &   & ...     &       ...     &      ...            &   & 72   &    0.06   &    0.632        \\[2pt]
$\alpha_{8.4}^{4.9}$   & \CIV\ blueshift & ...            &   ...     & ...                    &  &  ...  &    ...           &     ...             &   & 72   & 0.19        &   0.119          \\[2pt]
$\alpha_{8.4}^{4.9}$   & \CIII\ FWHM & 138  &   -0.07     &   0.433           &  & 68 &   0.07       &   0.553        &   & 70   & -0.011       &   0.923          \\[2pt]
$\alpha_{8.4}^{4.9}$   & \CIII\ blueshift & 138  &   -0.11    &   0.190            &  & 68 &    -0.12     &   0.331        &   & 70   & -0.03       &   0.786          \\[2pt]
$\alpha_{8.4}^{4.9}$   &  \MgII\ FWHM   &   68   &  -0.04   &    0.749      &   & 36   &   -0.22   &    0.199 &   & 32   &   -0.02    &    0.924   \\[2pt]
$\alpha_{8.4}^{4.9}$   & \MgII\ blueshift & 68   &   -0.16    &   0.193             &  & 36 &    -0.04     &   0.799       &   & 32    & -0.29       &   0.110          \\[2pt]
$\alpha_{8.4}^{4.9}$   & $\alpha_{uv}$ & 142  &   0.01     & 0.916          &  &   70  & -0.22     & 0.061         &   & 72   & -0.03     &   0.821          \\[2pt]
$\alpha_{8.4}^{4.9}$ ($\alpha_{fit}$) & $\lambda L_{2500}$    & 142 (145) & 0.17 (0.16) & 0.040 (0.057)     &  &   70 (71)  &  0.33 (0.26)     &\textbf{0.005} (0.028) &   & 72 (74) &  0.16 (0.21)    &   0.185 (0.072)         \\[2pt]
$\alpha_{8.4}^{4.9}$ ($\alpha_{fit}$) & $R^{*}$    & 144 (147)  &  -0.15 (-0.35)     & 0.070 (\textbf{${10^{-5}}$}) &  &   72 (73)  & -0.01 (-0.22)     & 0.948 (0.060) &   & 72 (74)   & -0.31 (-0.48)     & \textbf{0.009} (\textbf{${10^{-5}}$}) \\[2pt]
$\alpha_{8.4}^{4.9}$   &    $BI$/$AI$   &    ...   &     ...       &       ...           &  &   37/63  &  -0.32/-0.03    &   0.054/0.823       &   & ...     &   ...         &   ...                \\[2pt]
$\alpha_{8.4}^{4.9}$   & $v_{max}$/$v_{min}$&    ...        &     ...   &  ...  &  &   63/70  &  0.06/0.15     &   0.646/0.210        &   & ...     &   ...         &   ...                \\[2pt]
$BI$ ($AI$)                    & \CIII\ blueshift        &    ...  &     ...       &       ...                &  & 38 (63)  &   0.08 (0.10)    &   0.616 (0.436)        &   & ...     &   ...         &   ...                \\
$v_{max}$ ($v_{min}$)& \CIII\ blueshift       &    ...   &     ...      &       ...                &  & 63 (70)  &   0.01 (-0.02)  &   0.936 (0.870)        &   & ...     &   ...         &   ...                \\
$BI$ ($AI$)                     & \MgII\ blueshift       &    ...  &     ...       &       ...                &  & 17 (31)  & -0.27 (-0.06) &   0.286 (0.730)        &   & ...     &   ...         &   ...                \\
$v_{max}$ ($v_{min}$)& \MgII\ blueshift       &    ...   &     ...      &       ...                &  & 31 (38)  &  0.11 (0.10)  &   0.541 (0.542)        &   & ...     &   ...         &   ...                \\
$BI$ ($AI$)                       &  $R^{*}$          &    ...  &     ...       &       ...           &  &   39 (65)  &   0.05 (0.05)    &   0.776 (0.700)        &   & ...     &   ...         &   ...                \\
$v_{max}$ ($v_{min}$)  &  $R^{*}$         &    ...   &     ...      &       ...           &  &   65 (72)  &   -0.23 (-0.29)  &   0.065 (0.011)        &   & ...     &   ...         &   ...                \\
$BI$ ($AI$)                        &  \FeII                 &    ...   &     ...       &       ...          &  &   39 (65)  &   0.06 (-0.05)   &   0.724 (0.691)        &   & ...     &   ...         &   ...                \\
$v_{max}$ ($v_{min}$)  &  \FeII                 &    ...   &     ...       &       ...           &  &   65 (72) &   0.18 (0.02)    &   0.148 (0.870)        &   & ...     &   ...         &   ...                \\
$BI$ ($AI$)                        & $\lambda L_{2500}$ & ... & ...    &    ...             &  &   39 (65)  &  -0.06 (0.03)  &  0.737 (0.798)         &   &  ...    &  ...          &   ...                 \\
$v_{max}$ ($v_{min}$)  & $\lambda L_{2500}$ & ... & ...    &    ...              &  &   65 (72) &   0.32 (0.15)    &  \textbf{0.008} (0.213)          &   &  ...    &  ...          &   ...                 \\
$BI$ ($AI$)                       & $\alpha_{uv}$            & ... & ...    &    ...              &  &   39 (65)  &   0.09 (0.03) &  0.584 (0.827)          &   &  ...    &  ...          &   ...                 \\
$v_{max}$ ($v_{min}$)  & $\alpha_{uv}$            & ... & ...    &    ...              &  &   65 (72)  &  -0.144 (-0.01) &  0.252 (0.945)         &   &  ...    &  ...          &   ...                 \\
$BI$ ($AI$)                       & $F_{edd}$ & ... & ...    &    ...              &  &   17 (31)  &   0.13 (-0.01)  &  0.612 (0.975)          &   &  ...    &  ...          &   ...                 \\
$v_{max}$ ($v_{min}$)  & $F_{edd}$ & ... & ...    &    ...              &  &  31 (38)  &   -0.06 (0.00)   &  0.709 (0.990)          &   &  ...    &  ...          &   ...                 \\
$BI$ ($AI$)                       &        $M_{bh}$              & ... & ...    &    ...              &  &   17 (31)  &   -0.12 (0.05)  &  0.652 (0.794)          &   &  ...    &  ...          &   ...                 \\
$v_{max}$ ($v_{min}$)  &        $M_{bh}$              & ... & ...    &    ...              &  &  31 (38)  &   0.21 (0.05)   &  0.260 (0.747)          &   &  ...    &  ...          &   ...                 \\
$\alpha_{uv}$               &  $R^{*}$     & 146  & -0.09    &     0.291      &  &   72  &   -0.16   &   0.171        &   & 74    &  -0.02   &   0.858         \\
$F_{edd}$   &  $M_{bh}$         & 71    & -0.66 & \textbf{${10^{-10}}$} & & 38 & -0.72 & \textbf{${10^{-7}}$} & & 33 & -0.48 & $\textbf{0.006}$ \\                       
$M_{bh}$                      & $R^{*}$            &  71     & -0.19    & 0.110          &  &   38      &-0.11              &        0.515              &    &  33   & -0.28    &   0.113            \\
$M_{bh}$                      & $L_{r}$            &  71     & -0.05    & 0.684          &  &   38      &-0.11              &        0.508              &    &  33   & 0.02    &   0.913            \\
\CIV\ blueshift               &  \CIII\ blueshift &    ...    &    ...        &     ...             &  &   ...        &        ...            &          ...                    &    & 72      & 0.26     &  0.028       \\
\CIII\ blueshift                & \MgII\ blueshift &    70  &  -0.10     &  0.414        &  &  37       &  -0.01            &       0.944               &     & 33    &  -0.21    &  0.242      \\

\hline
   \end{tabular}
  \\
  {
  \raggedright
  Correlations involving $\alpha_{rad}$ were checked with both $\alpha_{8.4}^{4.8}$ and $\alpha_{fit}$; only values using $\alpha_{8.4}^{4.9}$ are shown unless the results were significantly different.  All correlations were examined using the Spearman rank correlation coefficient ($r_s$), and separately using all the objects in the sample, just the BAL quasars, and just the non-BAL quasars.  These results do not change significantly when imposing a SNR cut or when excluding LoBALs from the BAL analysis.\\
  }
\end{table}
\end{landscape}

\end{document}

%% file: commands.tex
\newcommand{\CIV}{C\,{\sc iv}}
\newcommand{\MgII}{Mg\,{\sc ii}}

\newcommand{\AlIII}{Al\,{\sc iii}}

\newcommand{\CIII}{C\,{\sc iii}]}

\newcommand{\HeII}{He\,{\sc ii}}

\newcommand{\FeII}{Fe\,{\sc ii}}

\newcommand{\NV}{N\,{\sc v}}

\newcommand{\OII}{O\,{\sc ii}}
\newcommand{\OIII}{[O\,{\sc iii}]}
\newcommand{\OIIIsemi}{O\,{\sc iii}]}

\newcommand{\SiIII}{Si\,{\sc iii}]}
\newcommand{\SiIV}{Si\,{\sc iv}}

\newcommand{\rmn}[1]{{\mathrm{#1}}}

\def\lsim{\lower0.3em\hbox{$\,\buildrel <\over\sim\,$}}
\def\gsim{\lower0.3em\hbox{$\,\buildrel >\over\sim\,$}}